\documentclass[fleqn,usenatbib]{mnras}

\usepackage[T1]{fontenc}
\usepackage{ae,aecompl}

\usepackage{graphicx}	
\usepackage{amsmath}	

\usepackage{amssymb}	
\usepackage{cuted}
\usepackage{lipsum}
\usepackage{bm}
\usepackage{lmodern}
\usepackage{pdflscape}
\usepackage{rotating}
\usepackage{adjustbox}
\usepackage{capt-of}
\graphicspath{{Figs/}}
\usepackage[dvipsnames]{xcolor}
\usepackage[normalem]{ulem}   

\usepackage{float}
\let\oldhref\href
\renewcommand{\href}[2]{\oldhref{#1}{\hbox{#2}}}

\usepackage{color}

\definecolor{orange}{rgb}{1,0.5,0}
\newcommand{\lastchange}[1]{\textcolor{black}{{#1}}}

\newcommand{\ra}[1]{\textcolor{orange}{{RA: #1}}}

\newcommand{\suppressed}[1]{}

\newcommand{\earlyplan}[1]{ } 

\definecolor{teal}{rgb}{0,0.5,0.5}

\newcommand{\HI}{\mbox{H{\scriptsize I}} } 
\newcommand{\mHI}{\mathrm{H{\scriptsize I}}} 
\newcommand{\Tsys}{T_\mathrm{sys}}      
\newcommand{\mK}{\mathrm{mK}}      
\newcommand{\tset}{T16DPA}
\newcommand{\cell}{$C_{\ell}$}
\newcommand{\cellcross}{$C^\times_{\ell}$}

\title[Tianlai low-z surveys]{The Tianlai dish array low-z surveys forecasts }

\author[O. Perdereau et al.]{Olivier Perdereau$^{1}$\thanks{E-mail: olivier.perdereau@ijclab.in2p3.fr},
R\'eza Ansari$^{1}$,
Albert Stebbins$^{2}$, 
Peter T. Timbie$^{3}$,
Xuelei Chen$^{4,5,6,7}$,
\newauthor
Fengquan Wu$^{4}$,
Jixia Li$^{4,5}$,
John P. Marriner$^{2}$,
Gregory S. Tucker$^{8}$,
Yanping Cong$^{4,5}$,
\newauthor
Santanu Das$^{14}$,
Yichao Li$^{7}$,
Yingfeng Liu$^{4,5}$,
Christophe Magneville$^{9}$, 
Jeffrey B. Peterson$^{10}$,
\newauthor
Anh Phan$^{3}$,
Lily Robinthal$^{3}$,
Shijie Sun$^{4,5}$,
Yougang Wang$^{4}$,
Yanlin Wu$^{3}$,
Yidong Xu$^{4}$,
\newauthor
Kaifeng Yu$^{4,5}$,
Zijie Yu$^{4,5}$,
Jiao Zhang$^{11}$,
Juyong Zhang$^{12}$,
Shifan Zuo$^{13}$
\newauthor
\\
$^{1}$Universit\'e Paris-Saclay, CNRS/IN2P3, IJCLab, 91405 Orsay, France\\
$^{2}$Fermi National Accelerator Laboratory, P.O. Box 500, Batavia IL 60510-5011, USA\\
$^{3}$Department of Physics, University of Wisconsin Madison, 1150 University Ave, Madison WI 53703, USA\\
$^{4}$National Astronomical Observatory, Chinese Academy of Sciences, 20A Datun Road, Beijing 100101, P. R. China\\
$^{5}$ University of Chinese Academy of Sciences Beijing 100049, P. R. China\\
$^{6}$Center of High Energy Physics, Peking University, Beijing 100871, P. R. China\\
$^{7}$ Department of Physics, College of Sciences, Northeastern University, Shenyang, Liaoning 110819, P. R. China\\
$^{8}$ Department of Physics, Brown University, 182 Hope St., Providence, RI 02912, USA\\
$^{9}$ CEA, DSM/IRFU, Centre d'Etudes de Saclay, 91191 Gif-sur-Yvette, France \\
$^{10}$Department of Physics, Carnegie Mellon University, 5000 Forbes Avenue, Pittsburgh, PA 15213, USA\\
$^{11}$ College of Physics and Electronic Engineering, Shanxi University, Taiyuan, Shanxi 030006, P. R. China \\
$^{12}$ Hangzhou Dianzi University, 115 Wenyi Rd., Hangzhou 310018, P. R. China\\
$^{13}$ Department of Astronomy and Tsinghua Center for Astrophysics, Tsinghua University, Beijing 100084, P.R.China \\
$^{14}$ Department of Physics, Imperial College London, Kensington, London SW7 2AZ, London, England
}
\date{Accepted XXX. Received YYY; in original form ZZZ}

\pubyear{2022}
\begin{document}
\maketitle
\begin{abstract}
We present the science case for surveys with the Tianlai dish array interferometer tuned to the $\left[ 1300, 1400 \right] \mathrm{MHz}$ frequency range.  Starting from a realistic generation of mock visibility data according to the survey strategy, we reconstruct a map of the sky and perform a foreground subtraction. We show that a survey of the North Celestial Polar cap  during a year of observing time and covering an area of $150 \, \mathrm{deg^2}$ would reach a sensitivity of $1.5-2 \, \mK$ per $1 \, \mathrm{MHz} \times 0.25^2 \, \mathrm{deg^2 }$ voxel
and be marginally impacted by mode-mixing. 
Tianlai would be able to detect a handful $(\sim 10)$ of nearby massive \HI clumps 
as well as a very strong cross-correlation signal
of 21\,cm intensity maps with the North Celestial Cap Survey optical galaxies. 
We have also studied the performance of a mid-latitude survey, covering $\sim 1500 \, \mathrm{deg^2}$ centered on a declination of $\delta=55^\circ$, 
which overlaps the Sloan Digital Sky Survey footprint. 
Despite a higher noise level for the mid-latitude survey, as well as significant distortions due to mode mixing, Tianlai would be able to detect a highly significant 
cross-correlation between the 21\,cm signal and the Sloan spectroscopic galaxy sample. Using the extragalactic signals from either or both of these surveys, 
it will be possible to assess the impact of calibration uncertainties, antenna pattern uncertainties, sources of noise, and mode mixing for future surveys requiring higher sensitivity.
\end{abstract}





\section{Introduction} 
\label{sec:introduction}
21\,cm Intensity Mapping (IM) is a promising technique to map the cosmological large scale distribution of matter 
through the observation of 21\,cm radio emission/absorption of neutral hydrogen gas (\HI), while not requiring the detection 
of individual sources \citep{2001JApA...22...21B, 2004MNRAS.355.1339B}  and has been largely explored in the context 
of the search for the EoR (Epoch of Reionisation) signal  \citep{2008PhRvD..78j3511P,Morales&Wyithe2010}. 
Subsequently, it was suggested that 21\,cm Intensity Mapping surveys could be used to constrain Dark Energy through 
the measurement of the BAO scale \citep{2008PhRvL.100i1303C, 2010ApJ...721..164S,Ansari2012}.
Such surveys require instruments with large instantaneous bandwidth
and field of view and several groups have built dense interferometric arrays to explore IM, such as CHIME \citep{2014SPIE.9145E..22B} 
or Tianlai  \citep{Chen2012}  in the last decade. 
Smaller instruments such as PAON4 \citep{2020MNRAS.493.2965A} and BMX \citep{2020SPIE11445E..7CO}
have also been built to explore specific technical aspects of these arrays, as well as transit mode operation and calibration. 
CHIME has proved to be a powerful fast radio burst and pulsar observation machine \citep{2021arXiv210604352T} and 
has motivated the design and construction of larger, dish-based, dense interferometric arrays such as  HIRAX \citep{HIRAX} 
and CHORD \citep{2019clrp.2020...28V}.
Intensity Mapping is also being used  by several large radio-interferometers, 
such as LOFAR \citep{LOFAR2013}, MWA \citep{MWA2013}, HERA \citep{HERA2017}, LWA \citep{LWA2018} and is also planned to be used in SKA-low \citep{2020MNRAS.494.4043M}
to search for signals from the EoR and the cosmic dawn, 
at redshifts above $z \gtrsim 10$.   

Tianlai is an international collaboration, led by NAOC, which has built and operates 
two radio-interferometers dedicated to 21\,cm Intensity Mapping since 2016 \citep{2018SPIE10708E..36D}. 
A first instrument is composed of three cylindrical reflectors, equipped with a total of 96 
dual-polarisation feeds \citep{Li2020a}  while the second instrument, the Tianlai Dish Pathfinder Array (hereafter T16DPA) 
features 16 on-axis dishes, 6 meter in diameter, equipped with dual-polarisation feeds, and arranged 
in a near-hexagonal configuration. The two instruments are located in a radio quiet site in Hongliuxia, Balikun county, 
in the Xinjiang autonomous region, in northwest China.  The two arrays have been observing in the frequency band
$\left[ 700,800 \right] \mathrm{MHz}$, corresponding to the redshift range $z \sim \left[  0.775,1.029 \right]$.
We recently reported on the various aspects of the operation and performance of the Tianlai Dish Pathfinder Array 
\citep{2021MNRAS.506.3455W}.   

Detection of the cosmological \HI signal and the ability of large instruments to constrain the $\Lambda$CDM model, specifically the
dark energy equation of state, through IM surveys covering the redshift range $z \lesssim 3-6$ has been extensively explored for 
large dedicated instruments \citep{2015ApJ...803...21B,Ansari2018}), or with existing or planned general purpose instruments such as SKA \citep{2017MNRAS.466.2736V} and FAST \citep{2017A&A...597A.136S}.


Table~\ref{tab:cross_correlations} gives parameters of measurements of 21cm cross-correlations, meaning 21\,cm induced correlation of low angular resolution radio maps with optical galaxy redshift surveys. Until recently cross-correlations have only been detected with large single dish radio telescopes. The first detection of cross-correlations with an interferometric array, an array specifically designed for hydrogen intensity mapping, has been made by CHIME.

\begin{table*}
    \caption{Measured cross-correlations of redshifted \HI emission with optical galaxy redshift surveys are listed in the first set of entries.  The last two entries describes the proposed surveys analyzed in this paper which are expected to also measure auto-correlations.  GBT is the Green Bank Telescope.  NCCSz is an ongoing redshift survey described in the text.  The significance of the detection could not easily be determined from \protect\cite{pen2009first}. \protect\cite{2022MNRAS.510.3495W} uses an extended version of the GBT observations of \protect\cite{Masui2013}.  Both \protect\cite{2022MNRAS.510.3495W} and \protect\cite{2022arXiv220201242C} correlate separately the same radio data with three different redshift samples.  The two Tianlai dish surveys require separate radio observations. 
    }
    \centering
    \begin{tabular}{c|c|ccccc}
\hline
redshift range & significance & radio telescope & optical redshift survey  & reference\\
\hline
   0-0.042  & -           & Parkes   & 6dF              & \cite{pen2009first}          \\
0.53-1.12   & 4$\sigma$   & GBT      & DEEP2            & \cite{2010Natur.466..463C}   \\
0.58-1      & 6$\sigma$   & GBT      & WiggleZ          & \cite{Masui2013}             \\
0.057-0.098 & 5.7$\sigma$ & Parkes   & 2df              & \cite{anderson2018}          \\
0.6-1       & 4.8$\sigma$ & GBT      & WiggleZ          & \cite{2022MNRAS.510.3495W}   \\
0.6-1       &   5$\sigma$ & GBT      & eBOSS-ELG        & \cite{2022MNRAS.510.3495W}   \\
0.6-1       & 4.4$\sigma$ & GBT      & eBOSS-LRG        & \cite{2022MNRAS.510.3495W}   \\
0.78-1.00   & 7.1$\sigma$ & CHIME    & eBOSS-LRG        & \cite{2022arXiv220201242C}  \\
0.78-1.10   & 5.7$\sigma$ & CHIME    & eBOSS-ELG        & \cite{2022arXiv220201242C}  \\
0.80-1.43   &11.1$\sigma$ & CHIME    & eBOSS-QSO        & \cite{2022arXiv220201242C}  \\
\hline
   0-0.068  & $\sim 40\sigma$    & Tianlai dish  & SDSS main sample     & this paper (forecasts)\\
   0-0.068  & $\sim 15 \sigma$     & Tianlai dish  & NCCSz          & this paper (forecasts)\\
\hline
    \end{tabular}
\label{tab:cross_correlations}
\end{table*}

The two Tianlai pathfinder instruments are also interferometric arrays designed for hydrogen intensity mapping, but are smaller than CHIME. 
The two, especially the dish array, are too small to be sensitive to the cosmological 
21\,cm signal around $z \sim 1$. In this paper, we study the extragalactic \HI signals that could be detected by the T16DPA 
by tuning its frequency band to very low redshifts ($z \lesssim 0.1$), through a detailed simulation of the reconstructed signal,
taking into account the  instrument response and survey strategy. Such an \HI signal within reach of the instrument would make it possible 
to precisely assess  the instrument and data analysis performance regarding key issues such as the gain, phase and bandpass calibrations, the 
impact of instrument noise, knowledge of beam and array configuration on the reconstructed 3D maps, and the 
level of residuals after foreground subtraction. 
Radio instruments are diffraction-limited, resulting in frequency-dependent beams, and for an interferometer, the set of angular wave modes sampled by a given baseline varies with frequency. `Mode mixing' refers broadly to the impact of this frequency-dependent instrument angular response on foreground subtraction.

We present an overview of the science targets of the Tianlai Dish Array low-redshift surveys in section \ref{sec:lowzsurveys}, while the simulation and analysis methods common to the different science cases are discussed in section \ref{sec:surv-sensitivity}, as well the expected survey sensitivities. Possible direct detection of nearby \lastchange{large \HI over-densities, corresponding to galaxies or group of galaxies, referred to as \HI clumps in this paper,}  is  presented in section \ref{sec:HIclumpsdetection}. The prospects for detecting large scale structures at low redshifts ($z \lesssim 0.1$) in cross-correlation with the SDSS and NCCS optical galaxy surveys is discussed in section \ref{sec:21cmxoptical}. Our findings are summarised and further discussed in the last section, \ref{sec:discussion}. 

\section{Low redshift surveys with Tianlai}
\label{sec:lowzsurveys}

The Tianlai dish array reflectors are equipped with feeds having a frequency bandwidth much larger than the instantaneous 100\,MHz bandwidth of the digitisation and correlator system. The instrument observation band is defined by the analog RF filters and the local oscillator frequency, which can be easily modified.
It is planned to tune the Tianlai Dish Pathfinder Array (T16DPA) frequency band to observe in the $\left[1330,1430\right] \mathrm{MHz}$ corresponding to the redshift ranges $0. \lesssim z \lesssim 0.068$.

In addition, the T16DPA dishes are fully steerable, which allows targeted observations, 
although in drift-scan mode. 

The North Celestial Polar cap  (NCP), accessible to the Tianlai Dish Array,  presents 
several advantages and is an optimal target to carry out
deep, high sensitivity observations, as suggested in \citet{2016MNRAS.461.1950Z}.  
A preliminary analysis of long duration observations of the NCP at $z \sim 1$ with  T16DPA has also been presented in \citet{2021MNRAS.506.3455W}.  

Low redshift surveys are considered as a path to prove the effectiveness of a dense interferometric dish array  using transit mode observations. For an Intensity Mapping survey to succeed, several instrumental and analysis 
challenges should be overcome, and requires in particular:  \\
- Precise determination of the instrument bandpass response \\
- Complex gain (amplitude and phase) calibration \\
- Understanding electronic and environmental induced noise behaviour\\ 
- Removing effects of cross-coupling between feeds and correlated noise \\
- Knowledge of array configuration, pointing errors \\
- Knowledge of dish antenna beam patterns and their impact on visibilities and
reconstructed 3D maps  \\
- Overall instrument and calibration stability \\  

As we shall discuss in detail below, there are low-z extragalactic \HI signals, with structuring in redshift similar 
to the cosmological LSS signals expected to be seen at higher redshifts, that will be within the T16DPA sensitivity reach. 
The observation of these extragalactic signals would enable Tianlai to assess quantitatively the impact of the above 
instrumental effects on the recovered signal. We shall also show that it would be possible to determine the residuals 
from foreground subtraction, the impact of individual antenna beams, and the bandpass response on these residuals. 

The advantage of observing at lower redshifts for T16DPA can be understood in two ways. On one hand, obviously, 
signals originating from individual extragalactic sources are much stronger at lower redshifts, say $z \sim 0.1-0.2$, than at redshifts 
$z \sim 1$, because the signal strength decreases as $d^{-2}_L(z)$, where  $d_L(z)$ is the luminosity distance to redshift $z$. However, one might argue that Intensity Mapping does not observe individual sources, but the aggregate 
emission from neutral hydrogen in voxels of $100-1000 \, \mathrm{Mpc}^3$ volume. Indeed, for a given setup, instrument angular resolution 
varies with redshift as $ (1+z)$, leading to the transverse voxel size evolving with $z$ as $ (1+z)^2 d_M(z)^2$, 
where $d_M$ stands for the transverse comoving distance with $d_L =(1+z) d_M$ and the angular diameter distance $d_A = d_M / (1+z)$  \citep{1999astro.ph..5116H}.  
So, ignoring the cosmological evolution of the source properties, the averaged intensity per voxel  would not vary with redshift. 
Nevertheless, considering the T16DPA angular resolution of $0.25^\circ - 0.5^\circ$, the voxel transverse size would range from $\sim 2 \,\mathrm{Mpc}$ at $z \sim 0.1$ to  $\sim 10 \,\mathrm{Mpc}$ at $z \sim 0.5$. The voxel size thus exceeds even the
cluster size at redshift 0.5, making direct detection of individual structures (galaxies, clusters) by T16PDA, quite unlikely beyond $z \gtrsim 0.1$, as will be shown in section \ref{sec:HIclumpsdetection}. 

What about statistical detection of LSS through the 3D map auto-correlation power spectrum? The LSS power spectrum changes slowly with redshift, contrary to distances. One might then expect that an IM instrument's ability to measure the LSS power spectrum would not change significantly with redshift. Unfortunately, the sensitivity to observe the cosmological LSS power spectrum decreases sharply as redshift increases, due to the way the radio interferometer's noise projects on sky. Indeed, a radio instrument, single dish or interferometer noise power spectrum, projected on sky as $P_\mathrm{noise}(k)$ scales as (see for example \citep{Ansari2012}, section 3.2): 
\begin{eqnarray}
  P_\mathrm{noise}(k,z)
  & \propto & d_M^2(z) \frac{c}{H(z)} (1+z)^2 \label{Eq.pnoise.A} 
\end{eqnarray}


This trend is due to the mapping from instrument coordinates, the two angles defining a direction
on sky and the frequency to a 3D position in a cosmological volume, but does not include any intrinsic noise level variation with frequency. We justify below this noise power $P_\mathrm{noise}$ evolution with redshift using a slightly different approach.

Let's consider brightness temperature sky maps $T_b(\alpha, \delta)$ with angular resolution $\delta \theta$ and frequency resolution $\delta \nu$. Instrument angular resolution $\delta \theta$ varies with wavelength $\delta\theta\propto\lambda/D_\mathrm{array}$ where $D_\mathrm{array}$ is the array spatial extent and $\lambda = c / \nu$ the observation wavelength.
Projecting such a map on a cosmological volume at redshift $z$, determined by the observation frequency $\nu$, we obtain voxels with transverse $a_\perp$ and radial $a_\parallel$ comoving dimensions, corresponding to a comoving volume $\delta V = a_\perp^2 \times a_\parallel$: 
\begin{eqnarray*}
  \nu & = & \frac{\nu_{21}}{1+z} \hspace{12mm} \nu_{21} = 1420.4 \, \mathrm{MHz} \\
  \delta \theta & = & (1+z) \delta \theta_0 \hspace{10mm}
      \delta \theta_0 = \delta \theta \left( \nu_{21} \right) \\
  a_\parallel & = & (1+z) \frac{c}{H(z) } \frac{\delta \nu}{\nu} = (1+z)^2 \frac{c}{H(z) } \frac{\delta \nu}{\nu_{21}} \\ 
  a_\perp & = & (1+z) d_A(z) \delta \theta = d_M(z) \delta \theta 
\end{eqnarray*}

The fluctuations  of the map  pixel's value due to instrumental noise, denoted $\sigma^2_T$ and characterised by the system temperature $\Tsys$ can be easily related to the noise power $P_\mathrm{noise}$.
The voxel dimensions $(a_\perp, a_\parallel)$ determines the maximum accessible wave numbers $(k_\perp, k_\parallel)$. Assuming white noise and ignoring damping due to
averaging over the voxel, we can write the Plancherel-Parseval identity:
\begin{eqnarray*}
  \sigma^2_T  & =  & \sum_{k_x, k_y, k_z} | F(k_x, k_y, k_z) |^2 \\
  \sigma^2_T  & = & P_\mathrm{noise} \iiint^{k^\mathrm{max}}_{-k^\mathrm{max}} \left( \frac{1}{2 \pi} \right)^3 d k_x d k_y d k_z \\
k_{\perp, \parallel}^\mathrm{max} = \frac{\pi}{a_{\perp, \parallel}}  & \rightarrow & 
P_\mathrm{noise} = \sigma^2_T \left( a_\perp^2 a_\parallel \right)  
\end{eqnarray*}
We would find a pessimistic $P_\mathrm{noise}$ redshift dependence if we apply the above formulae directly, compared to the one from equation \ref{Eq.pnoise.A}. Indeed, the array instantaneous field of view (FOV) increases with redshift like $\left(\delta \theta\right)^2$ as $(1+z)^2$, increasing the mapping speed. 
The per pixel noise would then decrease with redshift as $(1+z)^{-1}$. Taking this effect into account, we find a redshift dependence for the noise power spectrum identical 
to the one in equation \ref{Eq.pnoise.A}.
\begin{eqnarray}
  P_\mathrm{noise}(z)  & \simeq & (1+z)^2 d_M^2(z) \frac{c}{H(z) } \frac{\delta \nu}{\nu_{21} } \left( \delta \theta_0 \right)^2 \times  \sigma^2_T
\end{eqnarray}
where $\sigma_T$ denotes the per pixel noise level at $z=0$.  Despite this steep increase of the noise level with redshift, an intensity mapping survey at $z \gtrsim 1$ would be feasible using an array with several hundred dishes, thanks to the decrease of $\sigma^2_T$ or the noise power as the inverse of the number of antenna in the array. 

A survey of the NCP by T16DPA would be sensitive to spherical harmonics $Y_{\ell,m}$ of
order $\ell$ in the range $75 \lesssim \ell \lesssim 850$ at $\nu \sim 1400 \, \mathrm{MHz}$ (see section \ref{sec:surv-sensitivity}), 
corresponding to  angular scales $2 \pi/\ell$.

Taking into account evolution of the instrument angular scale range with redshift $(\ell \propto 1/(1+z))$,
we obtain the survey transverse wave number sensitivity range:
\begin{eqnarray}
  k_\perp (z) & = & \frac{\ell(z)}{ d_M(z)} \\
  \ell^\mathrm{min}(z=0) & \simeq & 75 \hspace{5mm}  \ell^\mathrm{max}(z=0)  \simeq  850 \\ 
  k_\perp^\mathrm{min,max} & = &  \frac{1}{(1+z) \, d_M(z)} \times \ell^\mathrm{min,max} (z=0)  
\end{eqnarray}

We have gathered in table \ref{tab-cosmo-scales} the voxel dimensions, and the accessible transverse $k_\perp$ range for a survey with angular scale sensitivities similar to T16DPA, map pixels with angular size $0.2^\circ$ at $\nu \sim 1400 \, \mathrm{MHz}$ and frequency resolution $1\, \mathrm{MHz}$. The projected noise level as a function of redshift is shown in figure \ref{fig-Pk-noise} as well as the accessible transverse $k_\perp$ range for a T16DPA survey.
The maximum value of the radial wave number
$ k_\parallel^\mathrm{max} $ is also listed assuming voxels with $\delta \nu = 1 \,\mathrm{MHz}$ resolution.
However, the T16DPA correlator computes visbilities with $\simeq 244 \,\mathrm{kHz}$ frequency 
resolution, so the survey could reach a maximum $ k_\parallel$ four times higher than the values listed in the table.
Unfortunately, all the foreground subtraction methods rely on the smoothness of synchrotron emission with frequency
and thus remove the signal modes with low  $k_\parallel$. The simulations we have carried out here suggest 
a low cut-off value $k_\parallel^\mathrm{min} \sim 0.15 k_\parallel^\mathrm{max}$ (see section \ref{subsec:noiselev-survey-sens}).

\begin{figure} 
\includegraphics[width=0.5\textwidth]{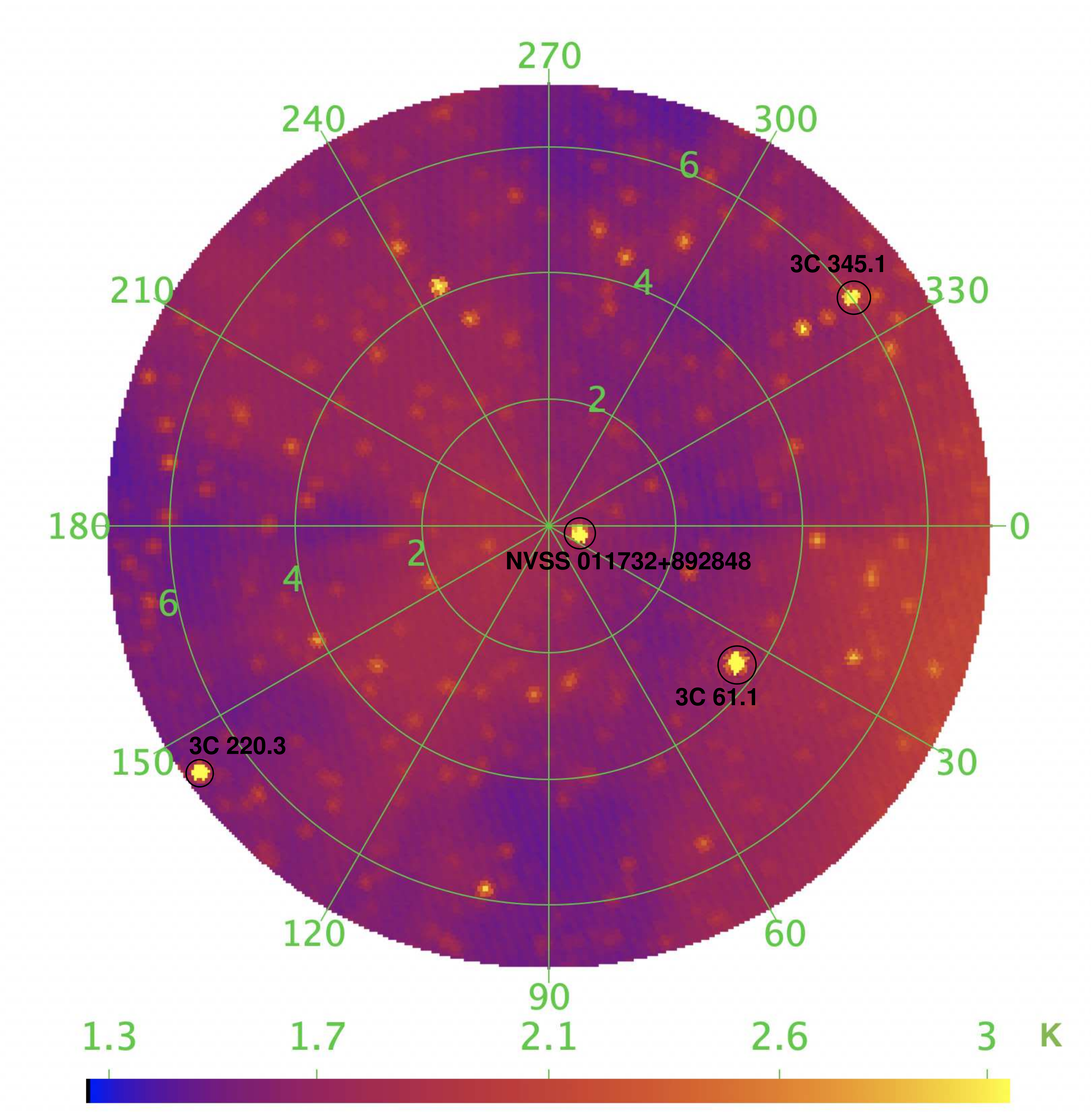}
\caption{Foreground map of a circular region of 7 deg. radius around the  NCP at 1350 MHz, smoothed with a $15 \,\mathrm{arcmin}$ resolution 
gaussian beam. The Haslam map of diffuse emission at 408 MHz as well as NVSS radio sources, extrapolated to 1350 MHz with a spectral index $\beta=-2$, have been co-added. The 4 brightest of these sources have been identified. The 
color scale corresponds to temperature in Kelvin.}
\label{fig-ncp-region-map} 
\end{figure}

\begin{figure} 
  \includegraphics[width=0.45\textwidth]{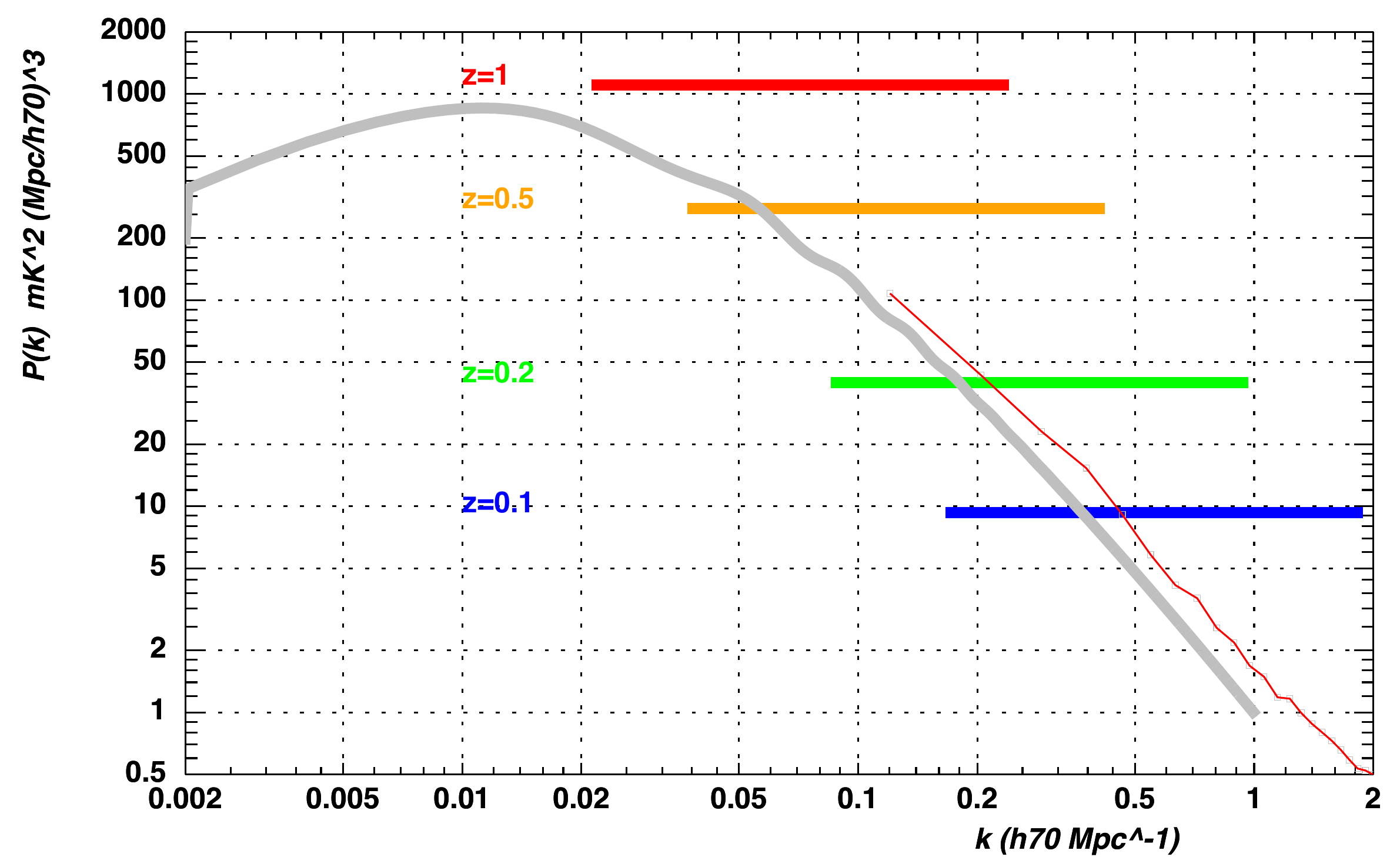} 
  \caption{
    Projected noise power spectrum $P_\mathrm{noise}(k)$ and the accessible transverse $k_\perp$
    range for a survey of the NCP region by T16DPA. The grey line show the expected linear 21 cm power spectrum at redshift $z=1$, while the red line shows the non linear power spectrum from simulations \citep{2018ApJ...866..135V} assuming a mean 21 cm brightness temperature $\bar{T}_{21} = 0.136 \,\mathrm{mK}$.
    \label{fig-Pk-noise}
    }
\end{figure}

Figure \ref{fig-ncp-region-map} shows the radio sky near the NCP (North Celestial Pole), as it appears at 1350 MHz, resulting from the combination of the Haslam map (dominated by synchrotron radiation) \citep{1981A&A...100..209H} and radio sources from the NVSS
catalog \citep{1998AJ....115.1693C}. We highlighted the brightest radio sources in this field. Their characteristics, retrieved from on-line archives such as  NED \footnote{\href{https://ned.ipac.caltech.edu/}{https://ned.ipac.caltech.edu/}} or with Simbad/Vizier\footnote{\href{http://simbad.u-strasbg.fr/simbad/}{http://simbad.u-strasbg.fr/simbad/}and \href{https://vizier.u-strasbg.fr/}{https://vizier.u-strasbg.fr/}, respectively}, are summarized in table \ref{tab:ncp_sources}. 
This field has also been observed by large scale radio interferometric arrays  such as  21CMA \citep{21cma_ncp} and LOFAR \citep{lofarNCP}. 

\begin{table*}
    \caption{Main characteristics of the brightest radio sources in the NCP field from figure \ref{fig-ncp-region-map}. (*) at 15 arcsec distance 
    \label{tab:ncp_sources}}

    \centering
    \begin{tabular}{c|c|ccccc}
\hline
NVSS id & other id & RA (hms) & Dec (dms) & 21cm flux  (Jy)       &  Nature & redshift \\
\hline
J011732+892848   & 6C 004713+891245 (*) & 1h17m32.82s & 89d28'48.7'' & 2.1 & &\\
J022248+861851 & 3C 61.1 & 2h22m35.046s & 86d19'6.17'' & 6 & Seyfert 2 gal. & 0.18781 \\
J093923+831526 & 3C 220.3 & 9h39m23.40s & 83d15'26.2'' & 2.95 & AGN & 0.685 \\
J213008+835730 & 3C 345.1  & 21h30m8.60s & 83d57'30.5'' & 1.8 & radio gal. & 0.865 \\
\hline
    \end{tabular}
\end{table*}


The visibility simulation and 3D map reconstruction is briefly described in section \ref{sec:surv-sensitivity}, as well as the
simple foreground subtraction methods we have used. We will  show in section \ref{sec:HIclumpsdetection} that it is possible to detect individual
galaxies or groups of galaxies at very low redshifts ($z \lesssim 0.05$) in the NCP region. We have also studied
the statistical detection of the LSS through cross-correlation with optical surveys, as discussed in section \ref{sec:21cmxoptical}. 
A mid latitude survey, covering a larger area, would be less sensitive due to higher noise level, but even more so due to much larger residuals from imperfect foreground subtraction, as discussed in  section \ref{sec:surv-sensitivity}. However, thanks to the
larger sky area, it would be possible to detect the cross-correlation signal with high statistical significance.

\begin{table}
  \caption{
  Listed as a function of redshift are the comoving distance, voxel dimensions, range of wavenumbers sampled and per voxel noise assuming $1\,\mathrm{MHz}$ pixels and per pixel noise of $\sigma_T^2=1\,\mathrm{mK}^2$.  Comoving lengths are in units of $\mathrm{Mpc/h_{70}}$, comoving wavenumbers in units of $\mathrm{h_{70} Mpc^{-1}}$ and white noise power $P_\mathrm{noise}$  in units $\mathrm{mK^2} / \left( \mathrm{Mpc/h_{70}} \right)^3$. See text for lower cut-off on $k_\parallel$ created by foreground subtraction. \label{tab-cosmo-scales} }
  \begin{tabular}{|c|c|cc|ccc|c|}
    \hline
    $z$ & $d_M$ & $a_{\perp}$ & $a_\parallel$ & $k_\perp^\mathrm{min}$ & $k_\perp^\mathrm{max}$ &  $k_\parallel^\mathrm{max}$ & $P_\mathrm{noise}$ \\
    \hline
    0.1 & 451 & 1.7 & 3.7  & 0.16 & 1.9 & 0.85 & 10 \\
    0.2 & 880 & 3.7 & 4.2  & 0.08 & 0.96 & 0.75 & 40 \\
    0.5 & 2028 & 10.6 & 5.5 & 0.037 & 0.42 & 0.57 & 280 \\
    1.0 & 3536 & 24.7 & 7.3 & 0.021 & 0.24 & 0.43 & 1100 \\
    2.0 & 5521 & 57.8 & 9.7 & 0.013 & 0.15 & 0.32 & 3600 \\
    \hline
  \end{tabular}

\end{table}

\earlyplan{  
\begin{itemize}
\item Explore also some aspects of component separation (foreground subtraction) 
\item Discuss the way instrument noise (radiometer equation) project on sky - when observations are discussed as cosmolgical power spectrum $P(k)$:
variations with reshift - Discuss also the accessible k-range (wave-number) - depending on the survey area and instrument configuration
($k_\perp k_\parallel$). 
\item Present the the three cosmological signals we might aim for : direct detection of HI clumps at very low redshifts $(z < 0.05)$, cross-corerlation with optical surveys 
and possible LSS detection in auto-correlation at $z \sim 0.4 $. 
\end{itemize}
}

\section{Survey sensitivity}
\label{sec:surv-sensitivity}
\subsection{Simulation and analysis pipeline}
\label{subsec:pipeline}
The JSkyMap \footnote{
  \href{https://gitlab.in2p3.fr/SCosmoTools/JSkyMap}{https://gitlab.in2p3.fr/SCosmoTools/JSkyMap} 
  (also check the wiki pages) }
package has been used for computing visibilities for the Tianlai dish array and the survey strategies
studied in this paper. The package also provides several tools for reconstructing maps from transit visibilities.
Here, we have used the m-mode visibility computation and map making tools, which operate in the spherical
harmonics space $Y_{\ell,m}$ as described in \citep{2016MNRAS.461.1950Z} and \citep{Shaw2014}.
The simulation and analysis pipeline includes several other C++ or python software modules, which handle 
the preparation of the input data, such as the generation of \HI sources from optical catalogs, foreground subtraction,
source detection, power spectrum computation and optical - radio cross-correlation computation.

The study presented here uses only intensity maps, ignoring polarisation. Foregrounds have been modeled
through the co-addition  of the diffuse synchrotron emission, represented
by the reprocessed Haslam map at 408 MHz \citep{2015MNRAS.451.4311R} and the radiosources from the NVSS 
catalog \citep{1998AJ....115.1693C}. In practice, for each simulated observation frequency, diffuse synchrotron emission 
and radio-sources have been extrapolated from their reference frequencies, (408 MHz and 1400 MHz) using a 
constant spectral index for the diffuse component $(\beta \sim -2 \ldots -2.5)$.  
All sources with flux larger than $0.05\, \mathrm{Jy}$ and $\delta > 15^\circ$ have been included 
in the simulation.  

The array configuration used in the simulations corresponds to the actual positions 
of the Tianlai antennas. We have used a frequency dependent single dish beam pattern $B(\theta)$, with azimuthal symmetry 
and modeled as an Airy disk with an effective dish diameter $D_\mathrm{eff} = 5.6 \,\mathrm{m}$. 
\begin{eqnarray}
B(\theta)  \propto  \left( \frac{2 J_1(x)}{x} \right)^2
\hspace{10mm} x = 2 \pi \frac{D_\mathrm{eff}}{2 \lambda} \sin \theta
\end{eqnarray}
where $J_1$ is the order one Bessel function of the first kind, $\lambda$ is the wavelength and $\theta$ is the angle with respect to the dish axis. 

There are 120 different baselines, excluding auto-correlations and ignoring polarisation. Visibilities have been computed with a right ascension or time sampling of $\delta \alpha = 30 \, \mathrm{s}$, well below the array angular resolution $0.25^\circ-0.5^\circ$, and we have used a $\delta \nu = 1 \,\mathrm{MHz}$ frequency resolution. 
Two surveys have been studied here, spanning a total duration of several months, up to a year. 
\begin{enumerate}
\item A survey of the NCP region with 4 constant declination scans $\delta = 90^\circ , 88^\circ, 86^\circ , 84^\circ$, and covering 
an area of about $100 \, \mathrm{deg^2}$ around the north pole. We have  used a fiducial area within 7 deg. from the north pole, $\delta > 83^\circ$, which would yield a surveyed area $\sim 150 \,\mathrm{deg^2}$. 
\suppressed{The simulated visibility data set for each 
\hspace{2mm} $ \sim 4 (\delta) \times 120 (\mathrm{visi})  \times 2800 (\mathrm{time}) \simeq 1.35 \times 10^6. $}

\item A survey in a mid-latitude area, covering a much larger portion of  sky, using 6 constant declination scans at 
$\delta = 49^\circ , 51^\circ , 53^\circ , 55^\circ , 57^\circ, 59^\circ$, covering a $12^\circ$ band in declination $48^\circ \leq \delta \leq 60^\circ$,
representing about $\sim 12 \%$ of the sky or $\sim 2500 \, \mathrm{deg^2}$. However, we have excluded a region in right ascension contaminated by the galactic plane and bright sources such as Cas A and Cyg A when computing noise power spectrum and mode mixing residuals. The fiducial area used, $40^\circ<\alpha<260^\circ$, represents about $1500  \,\mathrm{deg^2}$.  \suppressed{The visibility data set represents about $2\times 10^6$ time samples per frequency plane and for each simulated case.}
\end{enumerate}

The T16DPA system noise temperature has been determined to be $\Tsys \sim 80 \, \mathrm{K}$ \citep{2021MNRAS.506.3455W}. 
The simulations performed in this study have been carried out with a fiducial noise level of $5 \, \mathrm{mK}$ per $\delta t =30 \,\mathrm{s}$ 
visibility sample, and for a $\delta \nu = 1 \,\mathrm{MHz}$ frequency band. Such a noise level should 
indeed be reached for a single linear polarisation after  8.5 days spent on each constant declination scan, 
corresponding to a total integration time $t_{int} = 8.5 \times 30 = 255  \,\mathrm{s}$, per $\delta \alpha=30 \,\mathrm{s}$ RA sampled visibilities, 
leading to a per polarisation noise level : \\[0.5mm]
\hspace*{5mm} $\sigma_{V_{ij}} = \dfrac{\Tsys}{\sqrt{t_{int} \delta \nu}} = \dfrac{80 \, \mathrm{K}}{\sqrt{255} \times 10^3} \simeq 5 \, \mathrm{mK}.  $ \\[0.5mm] 

As shown in \cite{2021MNRAS.506.3455W}, the Tianlai dish array daytime data is contaminated by the Sun signal leaking into the far side lobes. It is therefore planned to use only nighttime data. 
Taking into account the $\sqrt{2}$ gain expected in the noise level when combining the two orthogonal  linear polarisation components, T16DPA should be able to reach a noise level of $5 \,\mK$ per RA visibility sample by surveying the NCP region during two periods of 
one month each, separated by 6 months, to get nighttime coverage of the full right ascension range.
A noise level of $2.5 \,\mathrm{mK}$ would also be reachable by observing the NCP area over a  year, spending a month on each declination. 
A similar noise level reduction could be achieved for the mid-latitude survey. However, as will be shown in section \ref{subsec:noiselev-survey-sens}, the survey sensitivity is limited by larger mode-mixing residuals in this case.

For each frequency, a spherical map is reconstructed through m-mode map making. 
In this method, the linear system of equations
relating visibilities to the input sky is solved in spherical harmonics $a_{\ell, m}$ space. A pseudo-inverse method is used in the JSkyMap package. 
The numerical stability of as well as the noise level are controlled through parameters $r_\mathrm{PSI}$, which is the ratio of the smallest to largest eigenvalues retained for each inversion, and $\lambda_\mathrm{PSI}$, the absolute threshold on the minimal 
eigenvalue. The values of these parameters have been set to the medium level of $r_\mathrm{PSI}=0.02$ and $\lambda_\mathrm{PSI}=0.001$ for the analysis presented here. 

\begin{figure} 
\includegraphics[width=0.5\textwidth]{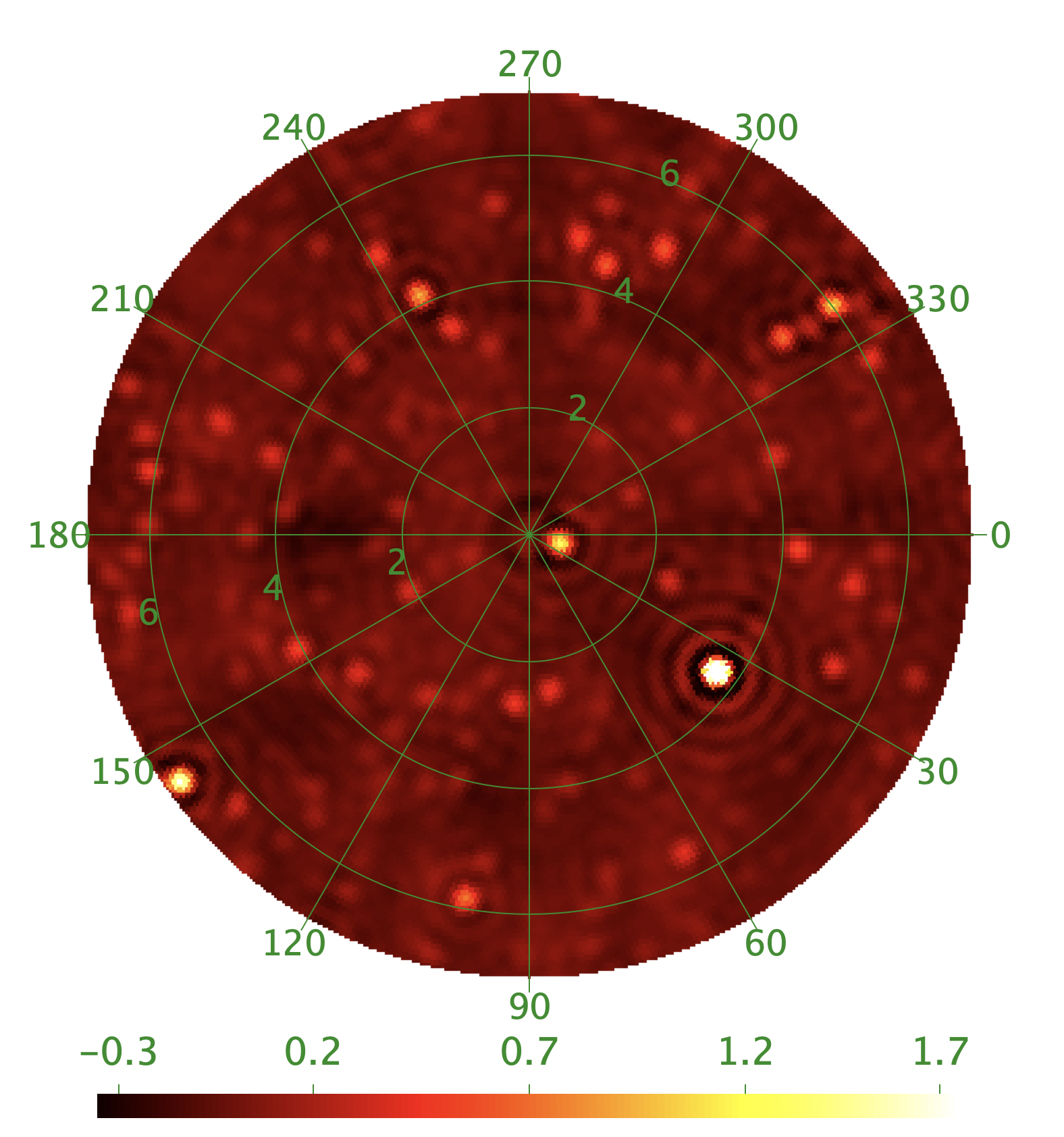}
\caption{Reconstructed map of the NCP region, as observed by T16DPA at  $f= 1350 \,\mathrm{MHz}$.
  This is the 
  $7^\circ$  radius area around $\delta = 90^\circ$, extracted from the reconstructed
  spherical map using m-mode map making and after $(\ell,m)$ space filtering. 
}
\label{fig-filtrecmap-ncp} 
\end{figure}

We have used spherical maps with a resolution of $5 \, \mathrm{arcmin}$,  although the array angular resolution is 
closer to $10-15 \, \mathrm{arcmin}$, as stated in section \ref{sec:lowzsurveys}. 
The reconstructed maps' pixels thus have  a certain degree of redundancy, with pixel to pixel noise values being correlated with each other for neighboring pixels. However 
these higher resolution maps presented a slight advantage for source detection and foreground removal. We have used the {\tt SphereThetaPhi} 
pixelisation scheme, which features almost square and equal area pixels along $\theta, \phi$ directions, implemented in the 
SOPHYA\footnote{SOPHYA C++ class library \href{http://www.sophya.org}{http://www.sophya.org} } library, 
instead of the more frequently used HEALPix scheme\footnote{\href{https://healpix.sourceforge.io/}{https://healpix.sourceforge.io/}}. This {\tt SphereThetaPhi}  scheme, belonging to the 
IGLOO pixelisations \citep{igloo_pix}, preserves to some extent the symmetry around a pixel located exactly at the pole $\theta=0$
and has also the advantage of being fully flexible in terms of angular resolution or pixel size. 

We also apply a filter in the spherical harmonics space $a_{\ell, m}$, before map reconstruction and foreground subtraction. 
The quality of the reconstruction degrades at the two ends of the T16DPA $\ell$ sensitivity range. At low $\ell$, this is explained 
by the absence of the autocorrelation signal, which is not used in map reconstruction, and the minimal baseline length, about 
$8.8 \,\mathrm{m}$, which limits the sensitivity below $\ell \lesssim 75$ for the NCP survey. At the other end, 
the noise level increases for $\ell \gtrsim 850$, which corresponds to the angular resolution of the array size for the NCP survey.
We have therefore smoothly damped $a_{\ell m}$ coefficients for $\ell \lesssim 75$ and $\ell \gtrsim 875$. A gaussian filter with $\sigma_\ell = 750$ has also been applied 
Moreover, all $m=0$ modes have been set to zero; this is intended to remove wiggles 
with near perfect azimuthal symmetry which appears due to the partial sky coverage combined with limited sensitivity range in $\ell$. 

\begin{figure}
\includegraphics[width=0.52\textwidth]{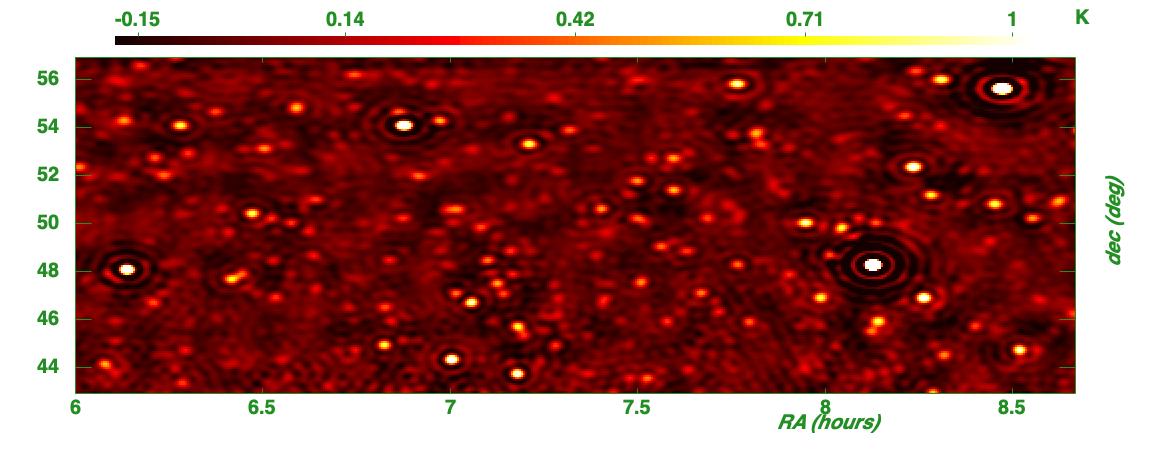}
\caption{Reconstructed map of the mid-latitude region,  after $(\ell,m)$ filtering, as observed by T16DPA at  $f= 1350 \,\mathrm{MHz}$. 
The patch of sky shown covers the declination range $43^\circ < \delta < 57^\circ$ and the right ascension range $90^\circ < \alpha < 130^\circ$, with $5 \, \mathrm{arcmin}$ pixel size.}
\label{fig-filtrec-map-midlat} 
\end{figure}

Figure \ref{fig-filtrecmap-ncp} shows an example of a reconstructed map, after $(\ell,m)$ space filtering at a frequency of 
$f= 1350 \,\mathrm{MHz}$. Sources present in the  {\em true} sky map (figure \ref{fig-ncp-region-map}),  as well as larger structures, 
are clearly visible, while the noise level ($2-4 \, \mathrm{mK}$) is too low to be noticeable. Some artefacts, such as rings  
around bright sources, can easily be seen and are due to  incomplete $(\ell,m)$ plane coverage and filtering. Similarly, a patch of reconstructed sky from a T16DPA survey of the mid-latitude area is shown in figure \ref{fig-filtrec-map-midlat}. This survey has a significantly higher noise level, 
compared to the NCP case, which is however not noticeable on this reconstructed map, 
as the brightest sources reach a few K.



\begin{figure} 
  \includegraphics[width=0.45\textwidth]{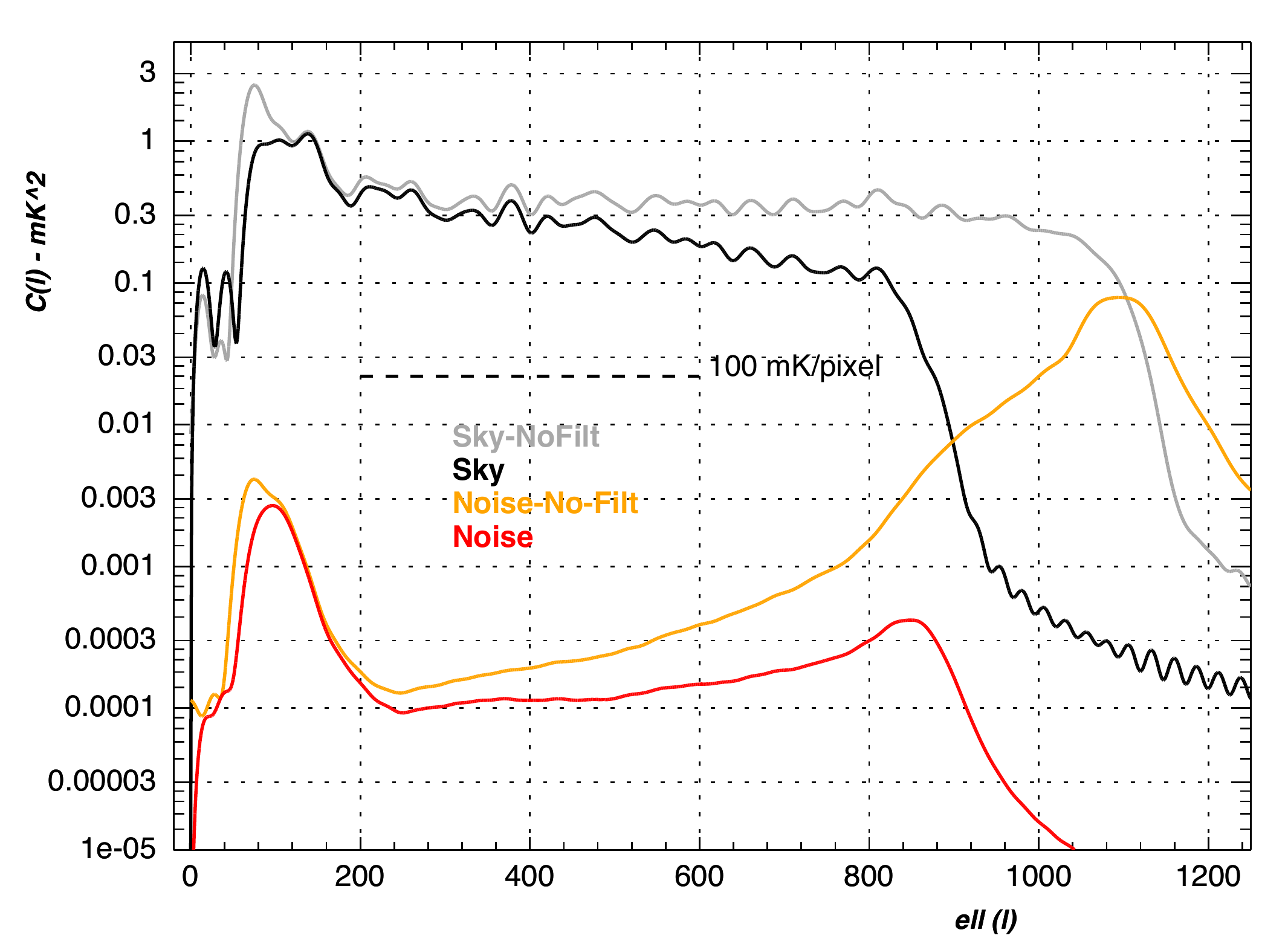} 
  \caption{Average angular power spectrum $C(\ell)$ from a data cube of 100 reconstructed maps covering an area with $7^\circ$ radius around the NCP and frequency range 1300-1400 MHz. The reconstructed sky power spectrum, as well as the noise power spectrum, are shown before (lighter colors) and after $(\ell,m)$ domain filtering (darker colors). The grey and black curves represent the sky power spectrum, before and after filtering, respectively. The orange and red curves show the power spectrum of maps reconstructed from noise only visibilities. }
  \label{fig-clskynoisefgs-ncp}
\end{figure}

The angular power spectra and noise of the reconstructed sky of the NCP region are shown in figure \ref{fig-clskynoisefgs-ncp}. 
The sky power spectrum is higher at larger angular scales, with an overall level of about $ \sim 1 \,\mathrm{K}$. The 
effects of the instrument and map-making response in the $\ell$ space are visible at the two ends, $\ell \lesssim 50$ and $\ell \gtrsim 1150$, of the unfiltered spectrum (grey curve). The additional effect of the $(\ell,m)$ domain filtering can clearly be seen by comparing the sky power spectrum before (grey curve) and after (black curve) this filtering. 
The projected noise angular power spectrum $C_\mathrm{noise}(\ell)$ 
is also shown on this figure. These $C_\mathrm{noise}(\ell)$ curves have been computed from maps reconstructed 
from white noise-only visibilities, with an RMS fluctuation level of $5 \,\mathrm{mK}$ per $\delta \alpha = 30\, \mathrm{s}$ 
visibility sample. As expected, the noise spectrum increases significantly  toward the high-$\ell$ end of the spectral sensitivity range, above $\ell \gtrsim 800$. This is due first to the decrease of the baselines' redundancy with $\ell$, and second, to the incomplete coverage of wave modes in the $(\ell,m)$ domain  at the high-$\ell$ end. The effect of $(\ell,m)$ filtering on the noise power spectrum $C_\mathrm{noise}(\ell)$ can be seen by comparing the orange curve, obtained before filtering, with the red curve,  after filtering.

\subsection{Foreground subtraction}
\label{subsec:foregroundsub}
Two simple foreground subtraction methods have been used here which exploit the 
smooth frequency dependence of the synchrotron-dominated foreground. The first method ({\bf P}) represents the synchrotron emission frequency dependence as a second degree polynomial
in frequency. The coefficients are determined for each direction through a linear $\chi^2$ fit to the measured temperatures, 
and the resulting fitted foreground $T^\mathrm{fgnd-P}_{\alpha, \delta}(\nu)$ is then subtracted from the 3D temperature map:
\begin{eqnarray}
T^\mathrm{fgnd-P}_{\alpha, \delta}(\nu) & = & A_{(\alpha, \delta)} \, \nu^2 + B_{(\alpha, \delta)} \, \nu + C_{(\alpha, \delta)} \\
T^\mathbf{P} ( \alpha, \delta, \nu) & = & T(\alpha, \delta, \nu) - T^\mathrm{fgnd-P}_{\alpha, \delta}(\nu)
\end{eqnarray}  
 
\begin{figure} 
  \includegraphics[width=0.48\textwidth]{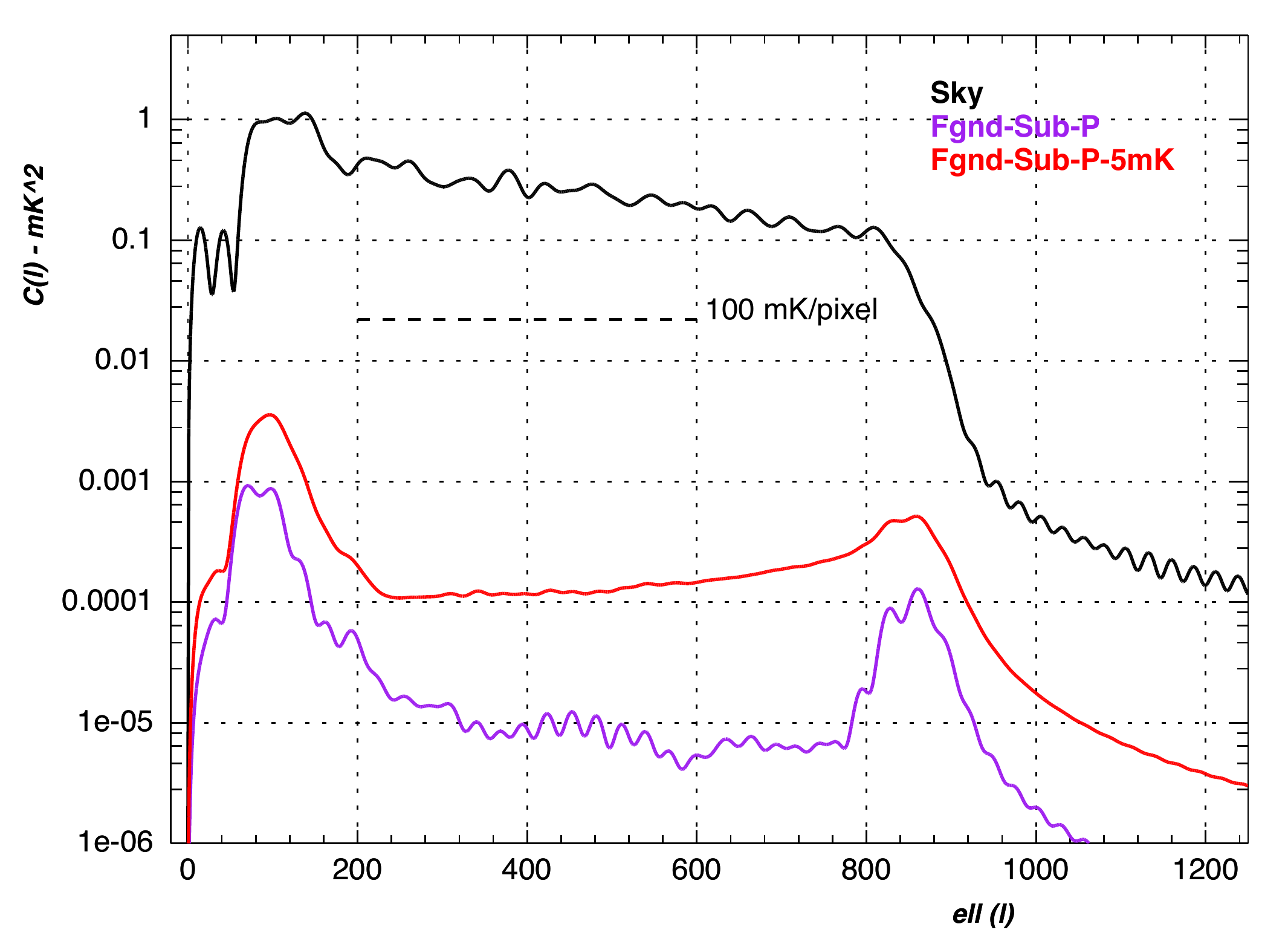}
  \caption{Similar spectra  as those shown  on figure \ref{fig-clskynoisefgs-ncp}, but now with foreground cleaning. The reconstructed sky power spectrum is shown in black, the residual after 
  foreground subtraction (polynomial fit-P) in purple without noise, and in red, with noise.
  }
  \label{fig-specskynoisefgs-ncp}
\end{figure}

The second method ({\bf DF}) is a difference filter along frequency. For each frequency $\nu$, 
we subtract the average of two nearby  frequencies $\nu_-, \nu_+$, with a specified frequency gap $\Delta \nu$;
$\nu_- = \nu - \Delta \nu$ and  $\nu_+ = \nu + \Delta \nu$ :
\begin{eqnarray}
T^\mathrm{fgnd-DF}_{\alpha, \delta}(\nu)  & = & \frac{1}{2} \left( T(\alpha, \delta, \nu_-) + T(\alpha, \delta, \nu_+ ) \right)  \\
T^\mathbf{DF} ( \alpha, \delta, \nu) & = & T(\alpha, \delta, \nu) - T^\mathrm{fgnd-DF}_{\alpha, \delta}(\nu)
\end{eqnarray}  
We have used $\Delta \nu = 2 \,\mathrm{MHz}$ throughout this paper. 

Figure \ref{fig-specskynoisefgs-ncp} presents the average angular power spectrum of the residual signal $C_\mathrm{res}(\ell)$ of
a set of 100 sky maps after foreground subtraction by the polynomial fit (P) method for the NCP survey. The maps are 
reconstructed from mock visibilities, 
for four constant declination scans at or near the NCP, and the $(\ell, m)$ plane filtering have been applied. 
Compared to the input sky angular power spectrum $C_\mathrm{sky}(\ell)$, shown as the black curve, one can see that the foreground angular power spectrum is suppressed by a factor $\gtrsim 20000$
for the polynomial subtracted foreground (P). This suppression factor reaches $\gtrsim 60000$ for the DF method (not shown). These values correspond to a factor $\sim 150$ (P) and $\sim 250$ (DF) damping in the amplitude for temperature fluctuations due to foregrounds. While this might not be sufficient for the direct detection of the cosmological 21 cm signal, the foreground residuals due to mode mixing and imperfect subtraction 
would be well below the instrumental noise level for the NCP survey by Tianlai. 

\begin{figure} 
  \includegraphics[width=0.48\textwidth]{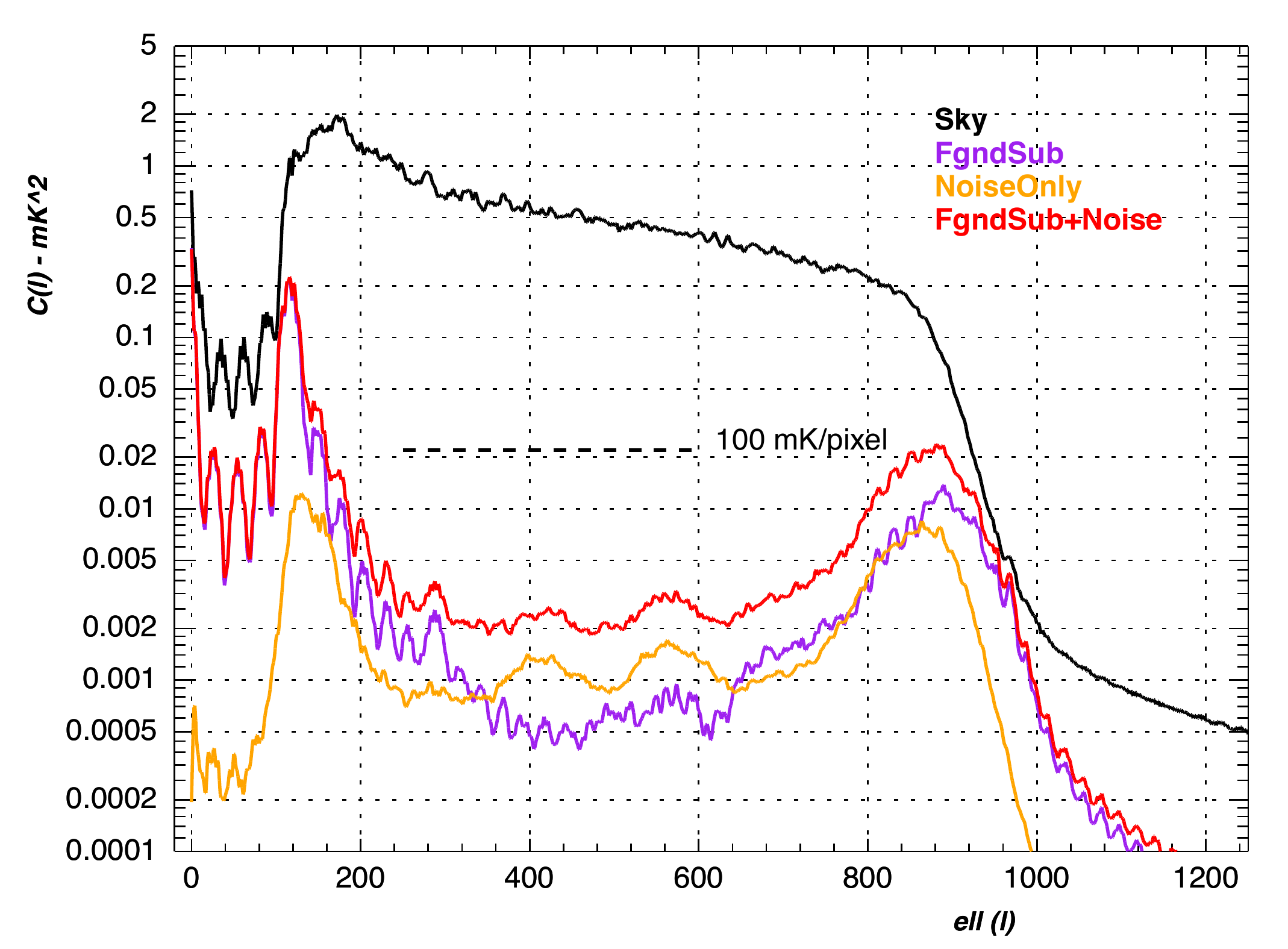} 
  \caption{Angular power spectra $C(\ell)$, for sky, noise and residuals after foreground subtraction 
    for the mid-latitude survey, computed from a set of three frequency maps at 1348, 1350, 1352 MHz. The differential filter along the frequency (DF) foreground subtraction method has been used here on $(\ell,m)$-space filtered maps. The black curve correspond to the reconstructed sky power spectrum, while the purple and red curves show the power spectra after foreground subtraction, without and with noise respectively. The orange curve corresponds to the power spectrum of the maps reconstructed from noise-only visibilities. } 
  \label{fig-specskynoisefgs-midlat}
\end{figure}

However, T16DPA becomes less efficient for mitigating  mode-mixing for a mid-latitude survey. The angular power spectrum of the residual after foreground subtraction using the DF method, for a mid-latitude survey, is shown in figure \ref{fig-specskynoisefgs-midlat}.
A set of three reconstructed maps at $f_{-2} = 1348 \,\mathrm{MHz}$, $f_0 =  1350 \,\mathrm{MHz}$ and $f_{+2} = 1352 \,\mathrm{MHz}$ 
have been used to compute these power spectra. 
A  fiducial area, representing $\sim 1500 \, \mathrm{deg^2}$, has also been used to exclude the right ascension ranges contaminated by the Galactic plane and Cas A and Cyg A. 
Comparing the black curve, which represents the reconstructed $C_\mathrm{sky}(\ell)$ 
of the diffuse synchrotron and radio sources, and the purple curve $C_\mathrm{res}(\ell)$, 
corresponding to the residuals after foreground subtraction, we see that the $C(\ell)$ power spectrum has been damped by a factor $\sim 1200$, or about $\sim 35$ for the amplitude of the temperature fluctuations. The residual after foreground subtraction reaches a level of $\sim 15\, \mathrm{mK}$, similar to the noise contribution.

The fact that this damping factor is 7 times lower in amplitude (about 50 times in the power spectrum) for the mid-latitude case compared to the NCP case is explained by  a higher level of mode-mixing for a mid-latitude survey by Tianlai, compared to the NCP case. Indeed, for observations toward the NCP, the sky orientation of the projected baselines changes with the sidereal rotation, improving the  map making performance 
in terms of individual mode reconstruction. 
The circular configuration of T16DPA was optimised for a good coverage of the angular sky modes or the $(u,v)$ plane, 
minimising the number of redundant baselines compared to a regular rectangular grid configuration, for example.
Although arrays with redundant baselines offer advantages for the gain and phase calibration, they would exhibit higher 
levels of mode mixing. For very large arrays, with several hundred or several thousand elements, a combination 
of redundant and non redundant baselines should be used to mitigate both mode mixing and calibration issues.

\subsection{Noise level and survey sensitivity}
\label{subsec:noiselev-survey-sens}

\begin{figure*} 
  \includegraphics[width=0.40\textwidth]{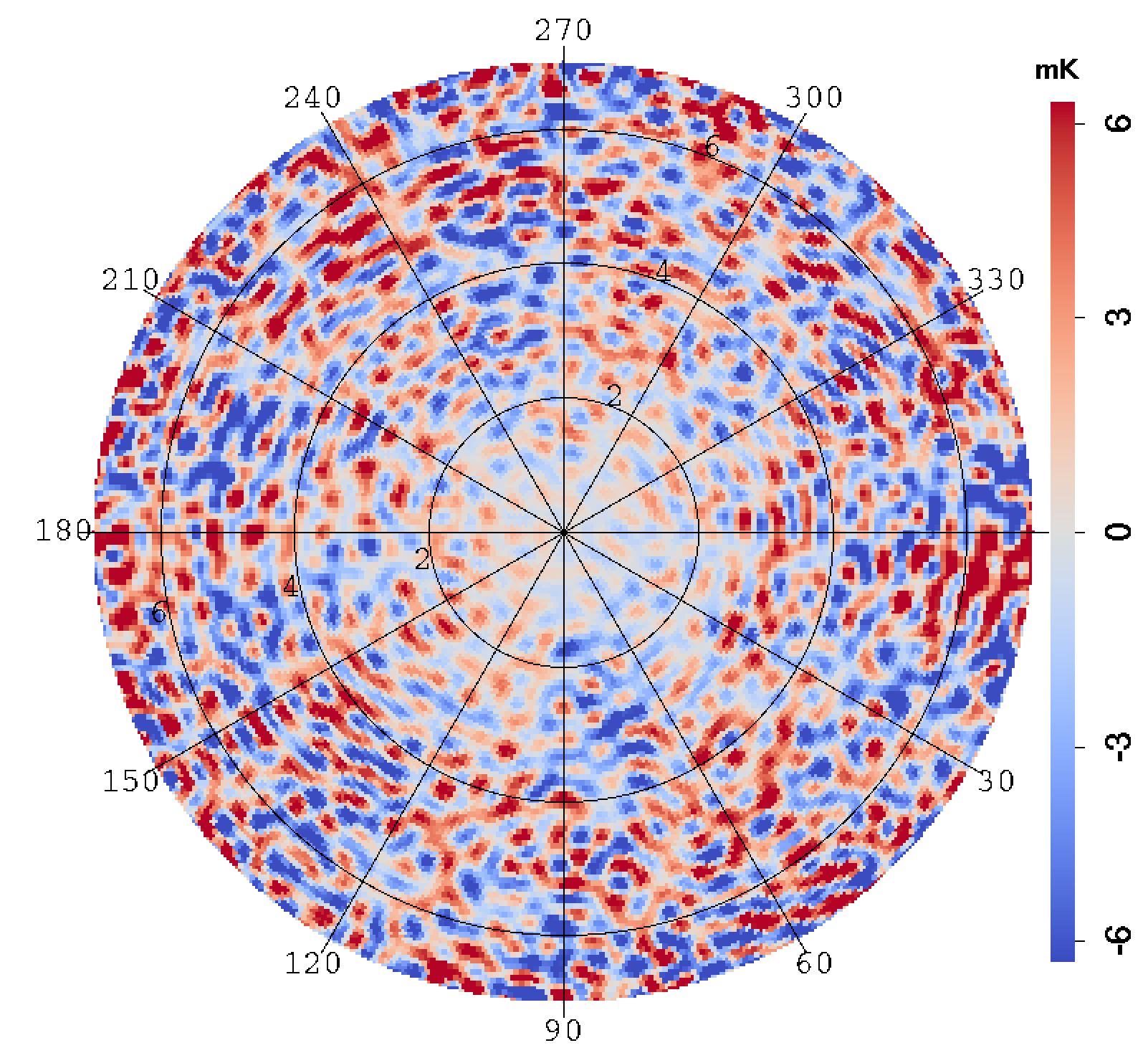} 
  \hspace{2mm} \includegraphics[width=0.46\textwidth]{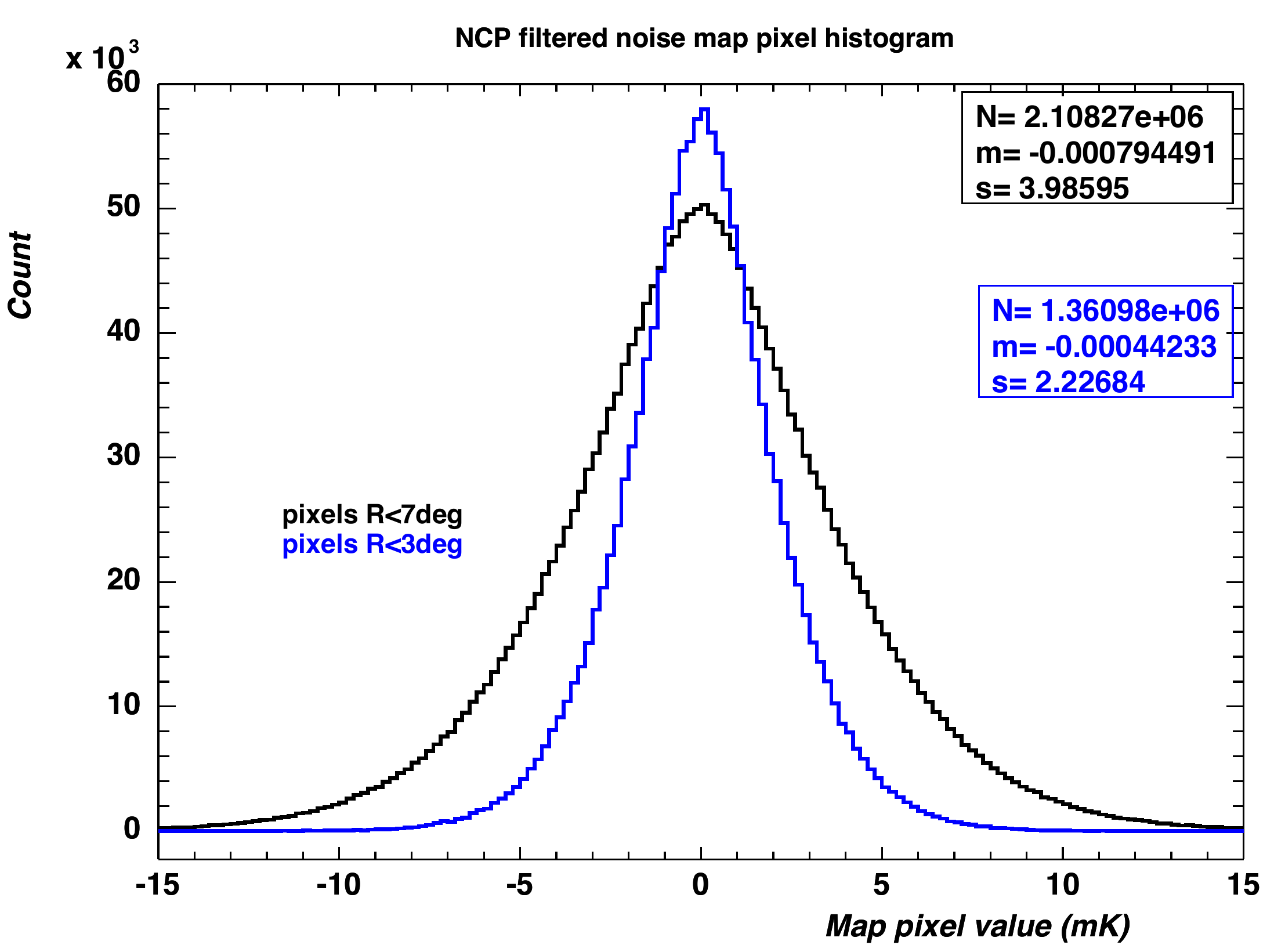}
  \caption{Left: Noise map after reconstruction with $(\ell,m)$
    filtering of the NCP region covering an area with 7 deg. radius at
    $f=1350 \mathrm{MHz}$ (map scale in mK). Right: Noise map pixel
    value distribution, in black for the full 7 deg. radius map around
  NCP, and blue, restricted to the central 3 deg. radius, covering $\sim 30 \mathrm{deg^2}$. 
  Note that restricted area corresponding to the blue histogram correspond to $\sim 18 \%$ of the full area; The blue histogram has been rescaled to enhance 
  the figure readability}
  \label{fig-noisemap-histo-ncp}
\end{figure*}

\begin{figure}
\includegraphics[width=0.52\textwidth]{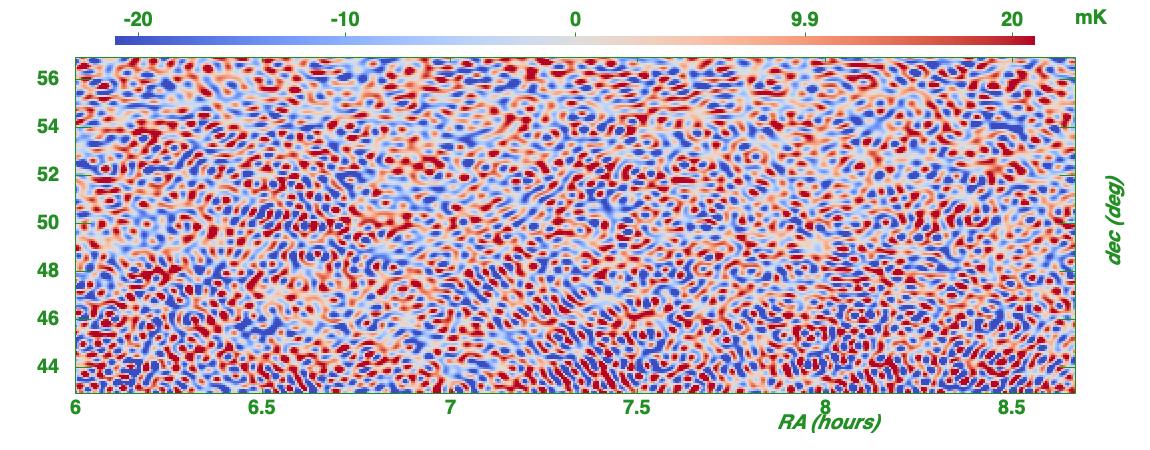}
\caption{Noise map of the mid-latitude region, corresponding to the reconstructed map shown in figure \ref{fig-filtrec-map-midlat}. }
\label{fig-filtnoise-map-midlat} 
\end{figure}

The left panel of figure \ref{fig-noisemap-histo-ncp} shows an example of a noise map for the Tianlai NCP survey, while the histogram of the corresponding pixel value distribution is shown on the right panel. 
The ($(5 \,\mathrm{arcmin})^2 \times 1 \,\mathrm{MHz}$) pixel to pixel temperature fluctuation 
is close to $4 \,\mathrm{mK}$, and even $2.2 \,\mathrm{mK}$ for the central $3^\circ$ radius area, assuming a $5 \,\mathrm{mK}$ noise level per $\delta \alpha = 30\, \mathrm{s}$ visibility sample. 
Figure \ref{fig-filtnoise-map-midlat} shows a similar noise map for the mid-latitude survey, with an RMS pixel fluctuation level of $\sim 16 \,\mathrm{mK}$. The noise level scales slightly faster than 
the square root of the ratio of the surveyed sky area, ($2500  \, \mathrm{deg^2} / 150 \,\mathrm{deg^2} \simeq 16$),
as the mid latitude survey discussed here requires 6 constant declination scans, hence 50\% more observing time.

As mentioned already, the maps with $5 \, \mathrm{arcmin}$ pixels used here have a higher resolution than 
the effective instrument and reconstruction angular resolution, which is limited to $\ell^\mathrm{max} \sim 850$ or $12 \, \mathrm{arcmin}$. The noise correlation between neighbouring  pixels is visible on the noise maps. 
The RMS fluctuation level decreases by a factor $1.5$ if maps with $0.25^\circ$ or $15 \, \mathrm{arcmin}$ pixels are used.
For the NCP survey the noise power spectrum $C_\mathrm{noise}(\ell)$ is nearly flat for $200 < \ell < 800$, and 15-40 times higher than the foreground subtraction residuals, as shown in figure \ref{fig-specskynoisefgs-ncp}. 
For the NCP region, Tianlai should be able to get down to the regime where the foreground residue is suppressed to a level below that of the noise
even for a deep survey that would reach $\sim 2 \,\mK$ per pixel noise level.

\begin{figure} 
  \includegraphics[width=0.48\textwidth]{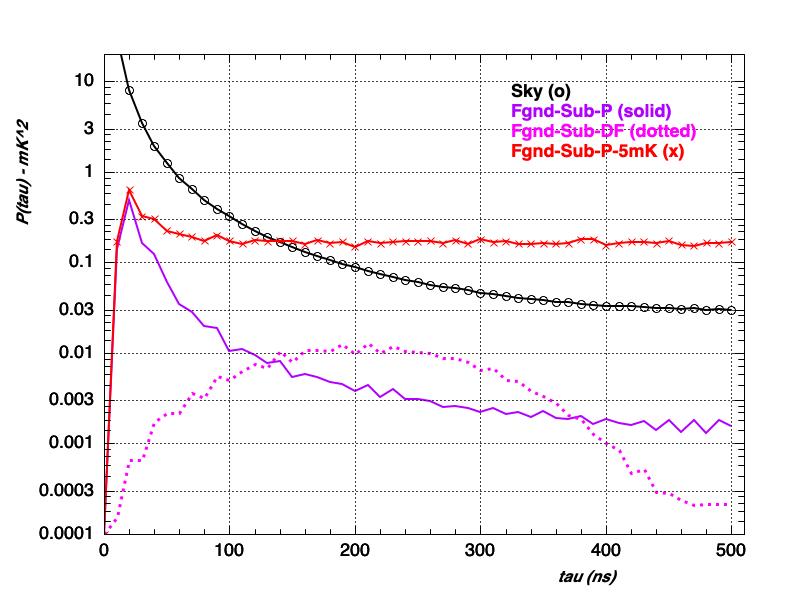}
  \caption{Average power spectrum along the frequency axis $P(\tau)$ from the a data cube of 100 reconstructed maps, with $(\ell,m)$ filtering of the NCP region. The horizontal axis correspond to the delay parameter $\tau$, ranging from 0 to $500 \,\mathrm{ns}$. 
  The reconstructed sky power spectrum is shown in black, while the purple (solid) and magenta (dotted) curves show the residuals after foreground subtraction, for the polynomial fit (P) and difference along the frequency (DF) methods respectively. The red curve correspond to the foreground subtracted map residuals, including noise and using Polynomial fit (P).
  }
  \label{fig-spec-tau-ncp}
\end{figure}

Figure \ref{fig-spec-tau-ncp} shows the average delay power spectra for the sky and residual after foreground subtraction 
$P(\tau)$ for the NCP survey.
These are the average of the spectra along the frequency direction, obtained through a radial Fast Fourier Transform (FFT), for each direction of the sky. 
The Fourier modes along the frequency \lastchange{are labeled as $\tau$ and correspond to a lag or delay time.}  
Given the 100 MHz bandwidth, with 100 frequency planes, the frequency modes or delay cover the range from $\tau_1 = 10\, \mathrm{ns}$ to $\tau_{50} = 500 \,\mathrm{ns}$. The black curve
represents the average reconstructed sky $P_\mathrm{sky}(\tau)$, with the power highly concentrated
at very low delay modes ($\tau \leq 20-30 \,\mathrm{ns}$), but with still significant power up to
$\tau \lesssim 100 \,\mathrm{ns}$. The effect of the two foreground subtraction methods and their $\tau-$response can be understood by looking at the shape of the average delay-spectrum
of the residual maps without noise (purple and magenta curves).  It can be seen that the polynomial foreground subtraction (P) suppresses 
delay-modes below $\tau \lesssim 50-60 \,\mathrm{ns}$, while the differential filter along frequency (DF)
can be considered as a band pass filter, removing $\tau \lesssim 100 \,\mathrm{ns}$ and
$\tau \gtrsim 400 \, \mathrm{ns}$. The (DF) method is more effective at removing foreground modes
at low delay, but leads to noisier maps. In addition to removing low-delay modes, the polynomial subtraction
method (P) damps the power $P(\tau)$ by a factor about $30$ for all modes above $100 \,\mathrm{ns}$.
The red curve correspond to the 
power spectrum $P(\tau)$ from maps after foreground subtraction (P method) with noise included, and it  
is noise dominated.

We have assumed a perfect knowledge of the instrument response, individual antenna beams pattern and perfect gain and phase calibration.  Discussion of the impact
of an imperfect knowledge of the instrument response on the survey performance is beyond
the scope of this paper. 
Preliminary studies suggest however that a phase calibration error between baselines, following a zero-mean normal distribution with an RMS of $7^\circ$, would result to a substantial increase of the foreground residual power spectrum, reaching a level 10 times higher than the one due to the instrument noise for the NCP case. 

\earlyplan{   left from the original paper plan - remove once done 
\begin{itemize}
\item Describe the visbility simulation and map reconstruction process
\item Describe and discuss the two foreground subtraction method used
  here   
\item per pixel noise level (visibility space and map space) 
\item Impact of imperfect calibration - (phase and amplitude calibration errors) 
\item illustrate for the NCP case, as well as lower latitude case 
\end{itemize}
}
 
\section{\HI clump detection}
\label{sec:HIclumpsdetection}


The aim of this analysis is to assess the number of direct detection of \HI clumps in a low-$z$ survey of either a mid-latitude band or a circular region around  the North Celestial Pole with T16DPA. We first estimate the \HI clumps detection efficiency as a function of their flux for the NCP and mid-latitude cases. 
In a second step we combine these detection efficiencies with 21cm flux derived from the ALFALFA  \HI clump mass function 
to determine the expected number of  detected clumps, assuming a spatial uniform random distribution, for the NCP and mid-latitude cases.

\subsection{\HI clumps flux detection thresholds}
To assess the detection efficiency for point-like \HI sources we have used a pipeline sharing most of the components 
described in section \ref{subsec:pipeline}. 
We simulate observations of the NCP and mid-lattitude surveys as described there, for only three frequencies : 1348, 1350 and 1342 MHz. 

To the generic astrophysical components (diffuse synchrotron Galactic emission , and continuum NVSS sources) we add, for the central frequency only, 
a set of uniformly distributed point-like sources of a given flux (in Jy). For each frequency we compute simulated visibilities, and then reconstruct sky maps 
and apply angular mode filter, as explained in \ref{subsec:pipeline}. 
The noise level per visibility sample ($30 \mathrm{s}$ integration time) used in the following is $5 \mathrm{mK}$.
In order to account for the impact of foregrounds, we have used the difference filter along the frequency ({\bf DF}) described in section \ref{subsec:noiselev-survey-sens}, which correspond to  subtracting from the central frequency map the average of the two outer frequency ones. 
Finally, we reproject the obtained sky map difference into rectangular (mid-latitude case) or square (NCP case) maps. 
An additional high-pass filter in the angular modes domain has been applied in the mid-latitude case to reduce foreground subtraction residuals. 

The final step of the pipeline is the source detection. We use a basic scheme based on  the {\tt DAOStarFinder} class from the {\tt photutils} Python package \citep{photutils_121}. 
Loose sphericity criteria for this source detector have been set, to compensate for remaining artefacts due to map reconstruction and foreground subtraction. 
We set the detection threshold, expressed as multiples of the map pixel-to-pixel RMS fluctuation level, to  7  and 10  for the NCP and mid-latitude
cases, respectively, to avoid spurious detections.  
The detection efficiency is determined by the number of detected sources within 2 pixels  of the simulated ones. 
In the NCP case, we simulated 5 sources over the 7  degrees circular observed region, but repeated this operation 20 times 
to reach a statistical accuracy of a few percent.  
In the mid-latitude case, the rectangular surveyed area is much larger, so that this iteration  is needed only 2 or 3 times.

\begin{figure*}
	\centering
	\begin{tabular}{cc}
		\includegraphics[width=.5\textwidth]{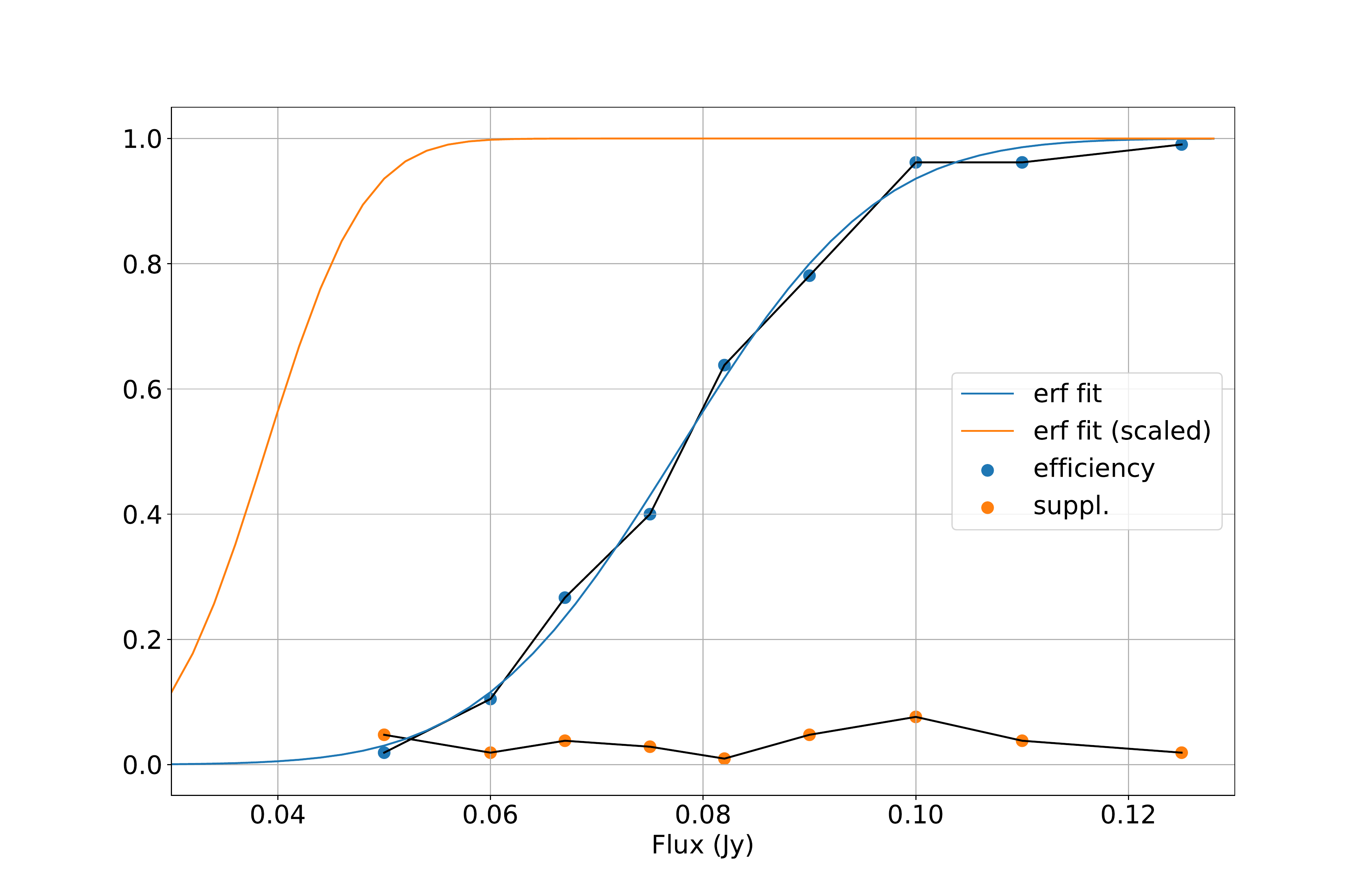}
		&
		\includegraphics[width=.5\textwidth]{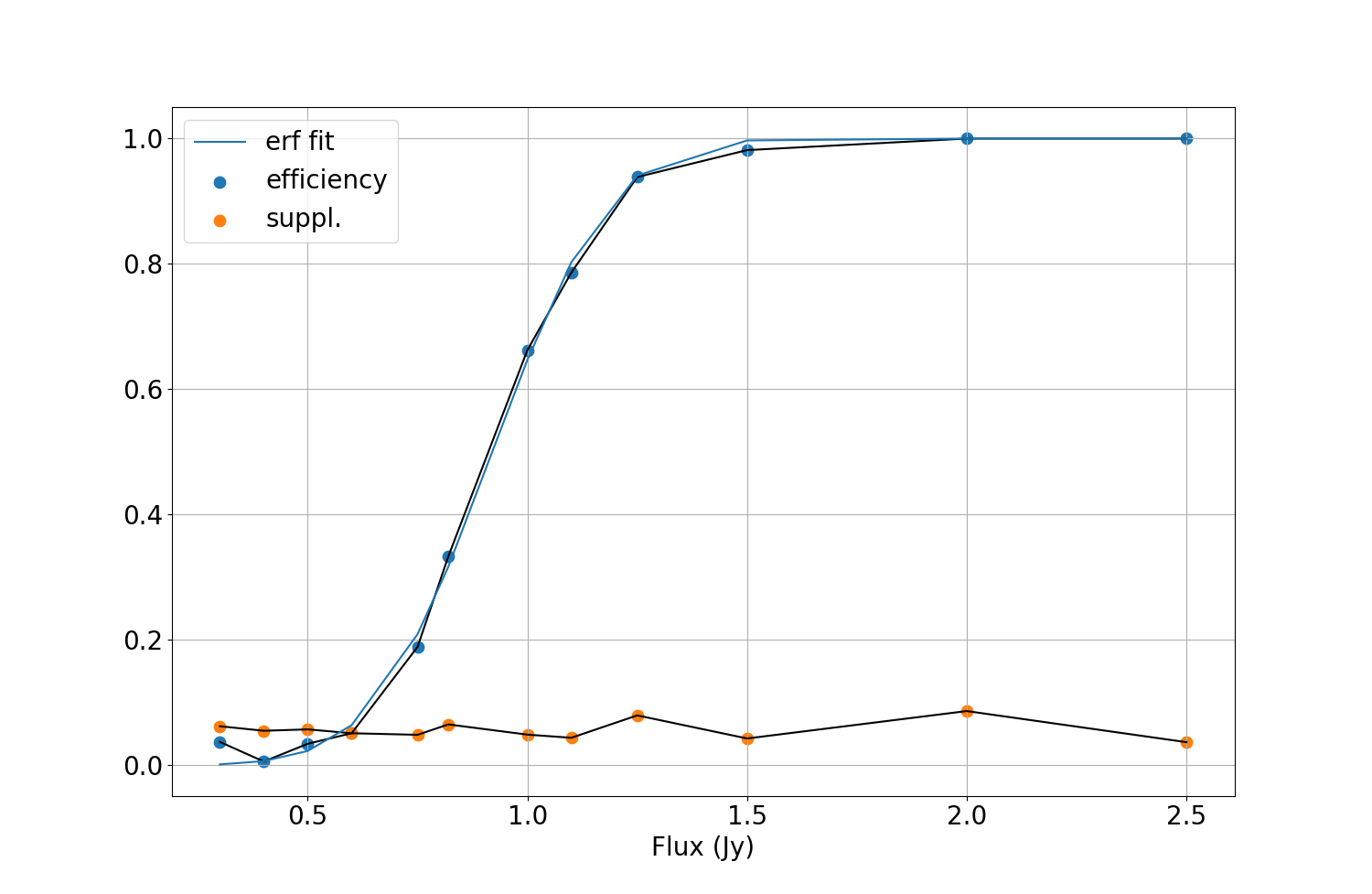}
		\end{tabular}
	\caption{HI clump detection efficiency as a function of flux, for the NCP (left-) and mid-latitude (right hand side) measured by  our simulations. On each part we represent as blue dots the efficiency measured at each simulated flux. the read dots correspond to the number of spurious detection (detections located farther than 2 pixels from the simulated clumps positions). The cyan curve is a fit of the efficiencies values with an error function. In the NCP case (left panel) we also indicate, in orange, what the efficiency function would become  if the noise per visibility sample was decreased by a factor $\sim$ 2 e.g. thanks to a longer integration time per declination. } 
		\label{fig:clump_det_effi}
\end{figure*}

The detection efficiencies we measure in the NCP and mid-latitude simulations are reported on figure~\ref{fig:clump_det_effi}. 
Thanks to the higher integration time per map pixel in the NCP case, these results show that the detection threshold $S_*^{th}$, 
defined as the flux limit with a detection efficiency $\geq 50\%$  is much lower in the NCP case 
than the mid-latitude one :  $S_*^{th} \simeq 0.08 \mathrm{Jy}$ for the NCP case, compared to $S_*^{th} \simeq 0.9 \mathrm{Jy}$ for the  
mid-latitude case. 
We have in each case fitted the source detection efficiency as a function of the flux to an error function $\mathrm{erf}(z)=\frac{2}{\sqrt{\pi}} \int_0^z e^{-t^2} dt$.  
These fitted functions have then been used in the computation of the expected number of 
\HI clump direct detections in the low-$z$ Tianlai surveys.  
We recall that the noise per visibility sample level of 5 mK used throughout this paper is 
a conservative estimate, as it could be achieved in only 3 months of observations (split into two halves separated by 6 month to avoid the sun-contaminated daytime), as explained in section \ref{subsec:pipeline}. We also indicate in figure \ref{fig:clump_det_effi} 
an estimation of the efficiency curve in this lower noise hypothesis in the NCP case, for which instrumental noise is the main limitation. 

\subsection{Number of expected \HI clumps observations}  
\label{sec:hi_clumps_numbers}
As shown in \cite{ingredients_21cmIM2018} (their section 4), in the redshift range considered in our analysis, most of the \HI mass lies in galaxies. 
We assume the \HI clumps population to have a random spatial distribution and to follow the characteristics measured using ALFALFA survey data by \cite{ALFALFAMassFunct}.
As shown in that paper, the \HI galaxy mass function - the number density of \HI galaxies in a logarithmic mass bins - is well described by a Schechter function :
\begin{equation}
\Phi(M) = \frac {dN_{\mHI}} {dV d \log_{10}(M)} = 
\log(10) \Phi^*\left( \frac{M}{M^*}\right)^{\alpha+1} \exp\left(-\frac{M}{M_*} \right)
\end{equation}
where $\Phi^*$ corresponds to the normalisation, $M^*$ the knee mass and $\alpha$, the low mass slope.  
\cite{ALFALFAMassFunct} fit these parameters using several subsets of the ALFALFA \HI source catalog  ; these results show some spatial dependence.  
We will retain here the parameters fitted with the whole dataset (ALFALFA 100\%) and its 'near' subset ($v_{CMB}<4000 \mathrm{km/s}$),  listed in Table \ref{tab:himfpars}.  The difference between the "full" and "near" parameter may give an indication of the systematics linked to the HI mass function.  ALFALFA also observed some variation of these parameters in different regions on the sky but we do consider different areas in this study, therefore we stick to this global variation with observed distance in the following. 

\begin{table}
\caption{HI mass function parameters determined   
	using either the whole ALFALFA dataset and its near subset \label{tab:himfpars}  }
    \begin{tabular}{|c|ccc|}    
	\hline 
	Dataset  & $\alpha$ & $m_*$ & $\phi_*$ \\
	\hline 
	Full & -1.25 $\pm$ 0.02 & 9.94 $\pm$ 0.01 & .0045 $\pm$ .0002 \\
	Near & -1.22 $\pm$ 0.02 & 9.76 $\pm$ 0.04 & .0062 $\pm$ .0005 
	\end{tabular}

\end{table}

We use the following expression to relate the \HI mass and the total 21cm flux $S_{21}$, also quoted in  \cite{ALFALFAMassFunct} :
\begin{equation}
\frac{M_{HI}}{M_\odot} = 2.356 \times 10^5 \left( \frac{D}{1 \mathrm{Mpc}} \right)^2 S_{21} 
\label{eq:s21_mhi}
\end{equation}
where $D$ is the source distance in Mpc and $S_{21}$ the integrated 21cm flux in Jy.km/s. 
For  each redshift value, we compute $D$ using fiducial cosmological parameters from  \cite{planck_cosmo_2015}.  Using this distance and assuming a $210 \mathrm{km/s}$
velocity width (corresponding to 1 MHz in the frequency domain) we can translate 
the flux limits  or detection efficiencies in Jy into \HI mass limits or detection efficiencies  at each redshift. 
The integral of the \HI mass function convolved with the detection efficiency gives the expected number density 
of clumps detectable by Tianlai at any given redshift. Integrating over the redshifts and 
taking into account volume element evolution with redshift, we obtain the 
expected total number of \HI clump detections. 

\begin{figure*}
	\centering
	\begin{tabular}{cc}
		\includegraphics[width=.5\textwidth]{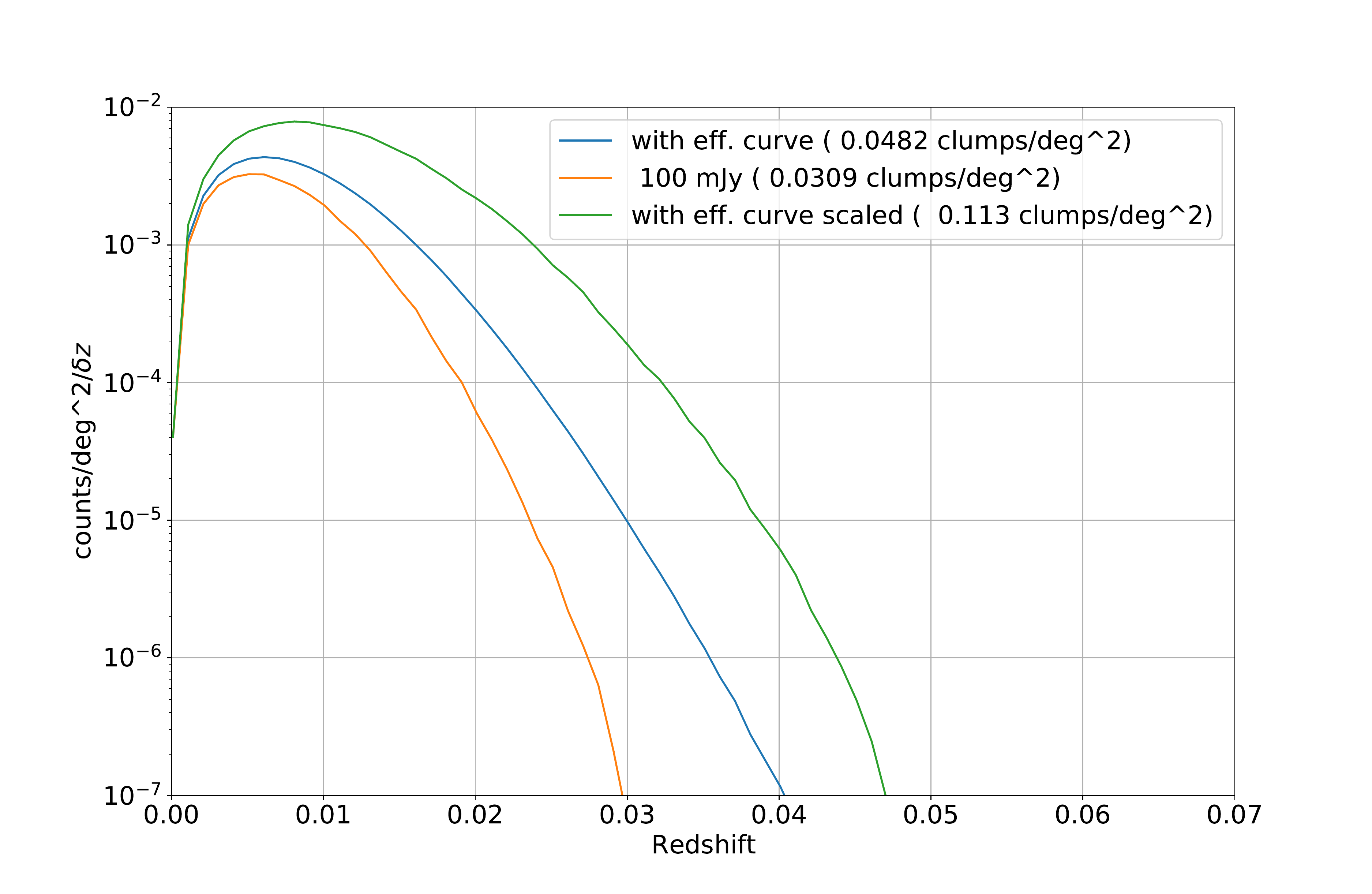}
		&
		\includegraphics[width=.5\textwidth]{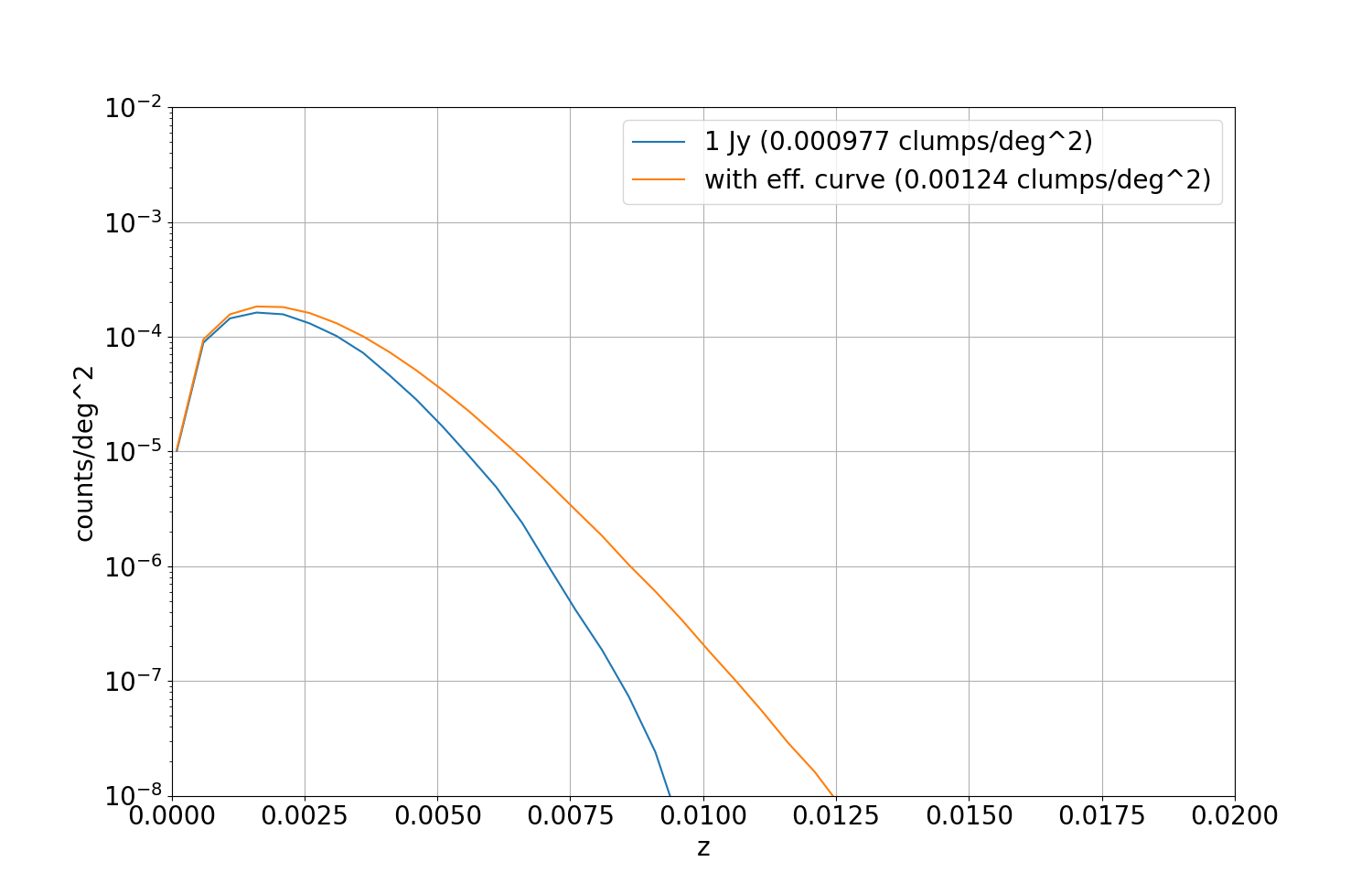}
	\end{tabular}
	\caption{Expected number of detectable \HI clumps as a function redshit, per square degree and per redshift bin $\delta z=0.001$,
 for the NCP (left panel) and mid-latitude (right panel) surveys. 
The total numbers of detections per square degree, integrated over redshift are reported in the caption of the figures.
Detection counts are shown using either a sharp detection threshold on the flux, in Jy (blue curves), or using the full shape of the detection efficiency curve (yellow). The green curve on the left panel (NCP) shows the number of detectable clumps with 
a lower noise $(\sim 2.5 \mK)$ on visibilities. }  
	\label{fig:clump_z_dist}
\end{figure*}

Figure \ref{fig:clump_z_dist} shows the expected number of \HI clump detections per square degree and 
per redshift bin ($\delta z = 0.001$) for the NCP  and mid-latitude cases as a function of redshift, 
assuming the \HI mass function parameters from ALFALFA full sample. 
The total number of detections per square degree are shown in this figure and also listed in table \ref{tab:hi_expt_dets}. 
As can be seen from  figure \ref{fig:clump_z_dist}, Tianlai would only be able to detect very nearby  
\HI galaxies, below $ z \lesssim .02$ for the NCP survey, and  $z \lesssim 0.005$ for the mid-latitude survey. 
The numbers vary slightly with the \HI mass function parameters used; if we use the {\it near} \HI mass function parameters from ALFALFA at these very low redshifts, the expected number of detection is even lower as the knee mass is somewhat lower in that case, 
as reported in table \ref{tab:hi_expt_dets}.

\begin{table}
	\caption{Number of expected \HI clump discoveries per square degree for the NCP and mid-latitude surveys, 
        for the two parametrizations of the \HI mass function given in table \ref{tab:himfpars}, 
        estimated with the fitted efficiency curves shown on figure \ref{fig:clump_det_effi}. 
        The sky area covered for each survey is given, as well as the total number of expected detections. 
        In the NCP case, we also indicate our estimations in the hypothesis of a lower noise per visibility sample 
        (or longer intergration time), using the corresponding efficiency curve.	\label{tab:hi_expt_dets} }

	\centering
	\begin{tabular}{|c|cccc|}
		
		\hline 
		Dataset  & NCP  & NCP & mid-lat. \\
	 	& & (low noise) & \\
		\hline 
		Surface ($\deg^2$) & 150 & 150 & 1500 \\
	clumps/$\deg^2$ (full MF) & .048  & .113 &  .0012 \\
	clumps/$\deg^2$ (near MF) & .035  & .083 &  .0009 \\
             N-clumps (full MF) & 7.2 & 16.95  & 1.8 \\
             N-clumps (near MF) & 5.25 & 12.45 & 1.35    
	\end{tabular}

\end{table}

In the mid-latitude case, the detection threshold is higher, hence a number of detection per redshift interval lower than in the NCP case is expected. This higher threshold is not totally compensated by the larger surface area covered by this survey, making the NCP survey the most promising  in terms of expected HI clump discovery rate. As mentioned in section \ref{sec:surv-sensitivity} 
a lower noise per visibility sample may be achieved by observing over longer period, less than a year for the full survey. In addition, 
the ${\bf DF}$ foreground subtraction method used for determining the source detection efficiency increases the noise level of the resulting 
difference map. The noise level which is the major limitation of the detection efficiency for the NCP case can be reduced by a factor about 2-3, 
combining lower visibility noise from the long survey duration and lower noise impact using the ({\bf P}) foreground subtraction.
A detection threshold $S_*^{th} \lesssim 0.05 \mathrm{Jy}$ could then be reached for the NCP survey, as indicated in figure \ref{fig:clump_det_effi}, leading to the expected number
of detectable clumps indicated in table \ref{tab:hi_expt_dets} (third columns), between 
12 and 17 in total, depending on the \HI MF used. 
 
\section{Cross-correlation with optical galaxy catalogs}
\label{sec:21cmxoptical}

\earlyplan{ discuss LSS in cross correlation with optical surveys , at mid-latitude, with SDSS , NCSS for NCP :
\begin{itemize}
\item discuss the possible scenarios : NCP, mid latitude , cross correlation with SDSS , 
\item show optical catalog redshift distribution. 
\item effect of incomplete spectroscopic catalogs 
\item effect of redshift errors 
\end{itemize}
}


In this section, we assess the prospects of detecting the cross-correlation signal between the intensity maps from the Tianlai low redshift surveys with 
optical galaxy catalogs:  SDSS \citep{SDSS-DR16} for the mid-latitude and NCCS \citep{GorbikovNCCS_I} 
for the polar cap survey.  
We first study the most straightforward case,  estimating the strength of the cross-correlation signal between a 
Tianlai mid-latitude low redshift survey and the overlapping portion of the SDSS catalog. 

\begin{figure*}
	\centering
	\begin{tabular}{cc}
		\includegraphics[width=.75\textwidth]{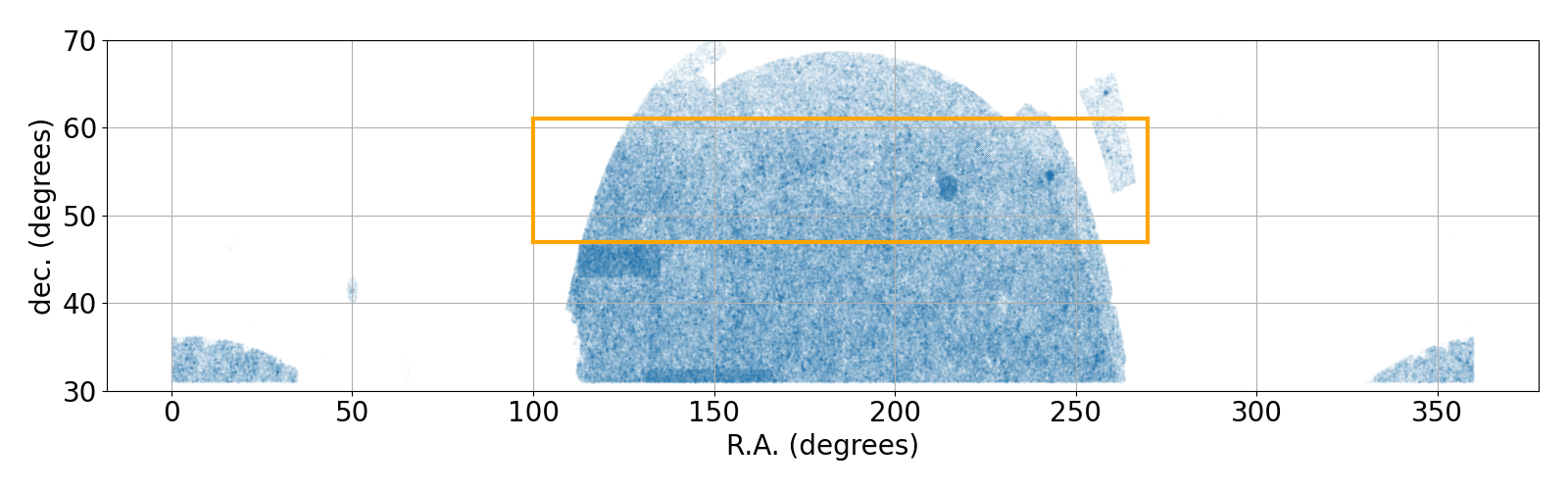}
		&
		\includegraphics[width=.25\textwidth,viewport=22mm 5mm 140mm 116mm,clip=true]{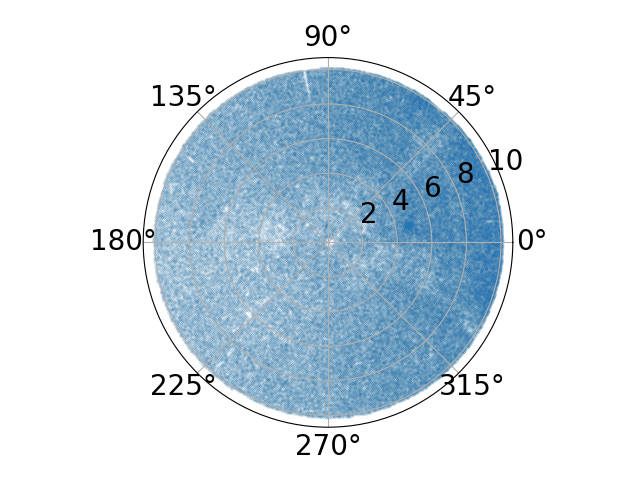}
	\end{tabular} 	
	\caption{Footprints of the SDSS (left) and NCCS (right) catalogs used in this paper. We selected galaxies above $\delta=30\deg$ in the SDSS catalogs and objects with  PESS (point vs extended source identification) score greater than 2 in the NCCS. The rectangular area outlined on the SDSS footprint is the area where cross-correlation with Tianlai low-$z$ simulated observations have been computed. We only plotted the positions of the NCCS sources with $V\leq17$. \label{fig:footprints} }
\end{figure*}

The SDSS catalog does not cover the NCP, but we are carrying out a spectroscopic survey using the WIYN telescope to obtain the spectroscopic 
redshifts for the brightest galaxies of the NCCS catalog.  
To evaluate the cross correlation signal for the NCP case, we  have used an artificial catalog built  
by rotating the coordinates of the objects in the SDSS catalog to overlap with our radio observations of the NCP. 
 The respective footprints of the SDSS and NCCS are shown on figure \ref{fig:footprints}. 

The selection criteria used to retrieve data from the SDSS DR16 server\footnote{\href{https://skyserver.sdss.org/dr16/en/tools/search/sql.aspx}{https://skyserver.sdss.org/dr16/en/tools/search/sql.aspx}} are given in Appendix \ref{sec:annex-sdss}.  
Starting from the optical galaxy catalog, we create a catalog of \HI sources using a two step procedure. We first derive a stellar mass 
from the optical galaxy properties, following  \cite{galactMstar}. The stellar mass is then converted into an \HI mass using the relations 
derived by \cite{alfalfa_mstar},  from the study of a combined ALFALFA-SDSS catalog. \HI emission parameters (flux and linewidth) are then derived from the \HI mass as explained in section \ref{sec:hi_clumps_numbers}. 
A more detailed description of the procedure for converting the optical catalog of galaxies into a list of 21cm source properties 
can be found in appendix \ref{sec:annex-photo2hi}.


The procedure used to determine the cross-correlation of Tianlai low-z observations with the SDSS or NCCS optical catalogs uses 
 the pipeline described in section \ref{subsec:pipeline}, with a few additional components:
\begin{enumerate} 
  \item A 21 cm source catalog is created from the SDSS galaxies with their redshifts or from the rotated SDSS catalog for the NCP case.
  \item Simulated visibilities that would be observed with the \tset\  setup is computed, combining signals from the different sky components: 
 diffuse synchrotron emission, radio sources, noise and redshifted 21 cm sources. Instrument noise is added to visibility samples as white noise. 
  \item Sky maps are reconstructed, independently for each frequency,  using  the m-mode decomposition method described in section \ref{subsec:pipeline}. A linear filter in spherical harmonic $(\ell,m)$ space is applied to compute spherical sky maps used in the next pipeline stages.
  \item The contribution of foreground emissions due to the Milky Way and radio sources is estimated and subtracted using either of the two approaches  presented in section \ref{subsec:foregroundsub}. 
  \item For the mid-latitude case, we project the filtered maps in an equatorial  band around the central latitude of the simulated observations, and select the relevant portion of this band for cross-correlation studies, as indicated on the left panel of figure \ref{fig:footprints}. 
  For the NCP case, the spherical maps are projected into square maps, for each frequency, using a Gnomonic or tangent plane projection centered on the North Celestial Pole.
  \item An optical source sky cube is also constructed from the optical catalog, using only the angular positions and the redshifts of the galaxies, ignoring the photometric information. All galaxies at the proper position and redshift interval 
    are included. 
  \item For each frequency, we compute the cross-correlation between the reconstructed radio sky map after reprojection and the corresponding 
    plane from the optical source cube. A correlation coefficient is computed, as well as a cross power spectrum, in the spherical harmonics domain, $C_\nu^\times(\ell)$,  
   for the mid-latitude case, and in the Fourier domain  $P_\nu^\times(k_\perp)$ for the NCP analysis.  
\end{enumerate}
%

\lastchange{It would of course be possible to characterize the cross-correlation 
through 3D power spectra $P^\times(k_\perp, k_\parallel)$. 
However, given that the signal fades away quickly with decreasing frequency at the very low redshift range considered here, and the different systematic effects affecting the radial $k_\parallel$ 
and transverse $k_\perp$ directions, we do not expect any advantage in using 
$P^\times(k_\perp, k_\parallel)$ instead of the set of $P_\nu^\times(k_\perp)$.}

The optical source cube to be correlated with the reconstructed sky cube has the same angular (5 arcmin pixels) and radial  resolutions (corresponding to 1 MHz frequency in the HI cube)
as the sky cube. Each galaxy in the optical catalog 
is assigned to a pixel in the cube. The frequency is determined from the source redshift, and the position in the plane from the angular coordinates of the galaxy. 
All galaxies have the same weight, equal to one, regardless of their photometric magnitudes. 
A gaussian smearing using the expected velocity width that we estimated as explained in appendix \ref{sec:annex-photo2hi} as FWHM along the frequency direction is then applied, 
as well as a 2D gaussian filter to each plane, with a  fixed angular width
$\sigma_\perp =10 \, \mathrm{arcmin}$.

We also build a series of randomized optical source cubes, called shuffles to determine the level of residual cross correlation signal, due to imperfect foreground subtraction and instrumental noise. 
The source angular positions and redshifts are shuffled independently from each other to generate these. We expect a null correlation of the reconstructed sky maps with these shuffled cubes. 
A hundred such  cubes have been built and correlated with 
the reconstructed sky cube to estimate the cross correlation signal dispersion. 
  

We have ignored the photometric information in building the optical source cubes in order to minimise the dependence of the cross-correlation signal strength with respect to the detail of the underlying model  relating the \HI properties to the optical magnitudes, and to avoid the need for generating multiple copies of the expected visibilities to take into account the dispersion in $M_{HI} = f (\mathrm{mag-}u,g,r,i,z)$ relation. The amplitude of the optical source cube is thus arbitrary, hence the normalisation of correlation spectra. Their comparison with each other, or when including or not different sky components remains however relevant. 

We ran the pipeline and analysed the results for two noise configurations, with or without $5 \mathrm{mK}$ noise added to visibility samples, and different combinations of sky components:
\begin{itemize} 
	\item[-] continuum sources only (Haslam-based synchrotron map and NVSS sources)
	\item[-] \HI simulated sources only
	\item[-]  all components, i.e. combining diffuse synchrotron, continuum radio sources, \HI sources and noise
\end{itemize}

\subsection{Mid-latitude survey cross-correlation with the SDSS catalog}
\label{sec:midlat_xcor}
We ran this pipeline for all  1 MHz frequency \lastchange{shells} between 1260 and 1410 MHz. We computed auto- and cross correlation \cell\  between 
the reconstructed sky cube and the source cube, built from the optical catalog for each frequency plane.  The cross correlation is also computed with each of the shuffled source cubes. 
One expects the cross power spectra between the randomized  source cube and the simulated maps to be null in average, 
and that the dispersion around their average will give an estimate of the uncertainty in the computed cross-correlation coefficient 
or power spectrum. 
The significance of the correlation in a single frequency bin is low, so we combine the results into 
three frequency (or redshift) intervals as described in Table \ref{tab:freq_z_bins}.

\begin{table}	
	\caption{Frequency bins used for the mid-latitude Tianlai-SDSS cross-correlation  analyses.\label{tab:freq_z_bins} }  
	\centering
	\begin{tabular}{|c|cc|c|}
		\hline 
		Bin   &  $\nu_{\mathrm{min}}$ (GHz) & $\nu_{\mathrm{max}}$ (GHz) & $z_{\mathrm{center}}$ \\
		\hline
		1 & 1260  & 1310 & 0.096 \\
		2 & 1310  & 1360 & 0.061 \\
		3 & 1360  & 1410 & 0.025 \\
	\end{tabular}
\end{table}

\begin{figure}
	\centering
	\includegraphics[width=.5\textwidth]{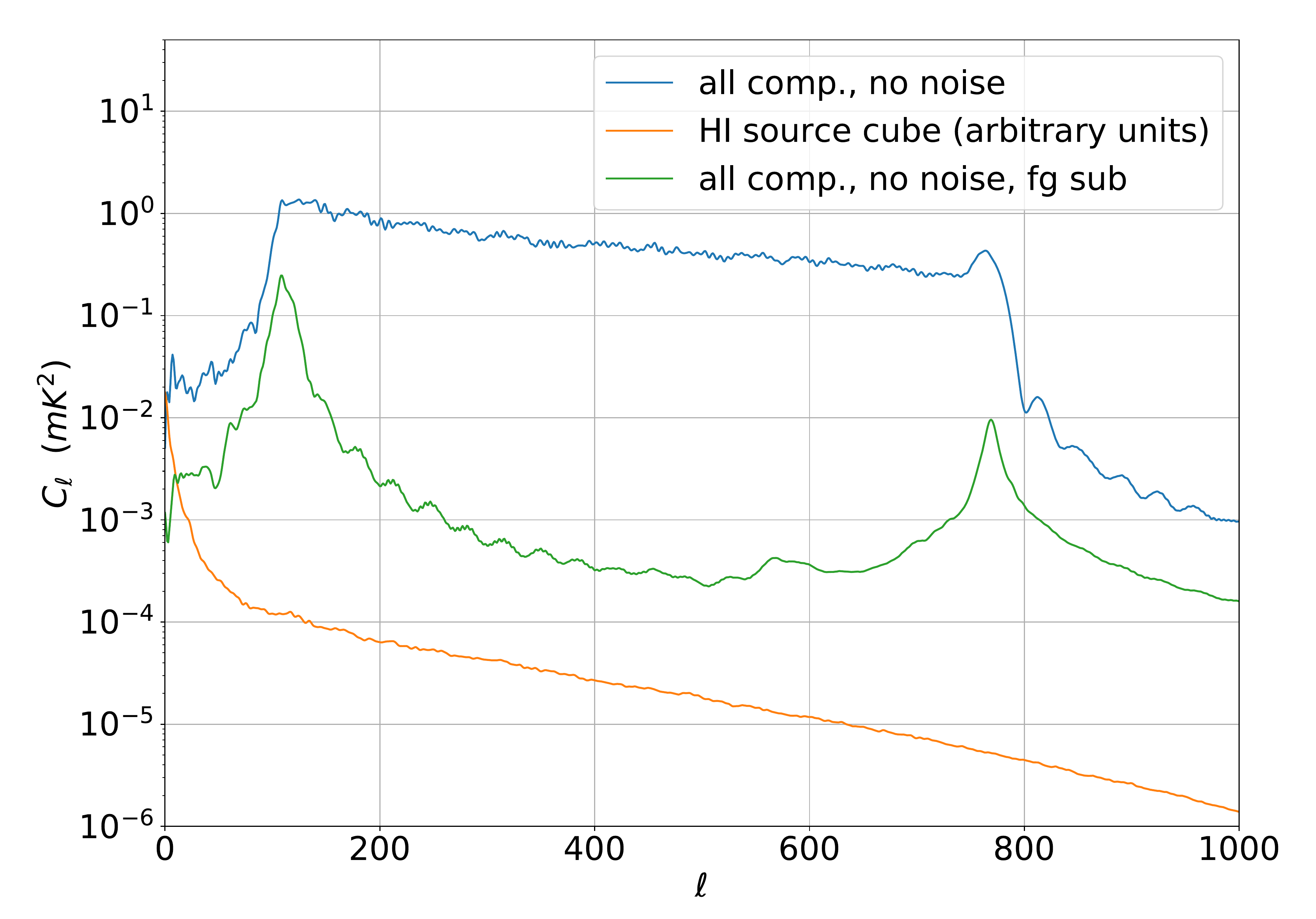}
	\caption{The averages of the auto-correlation spectra (\cell) for frequencies between 1350 and 1375 MHz. 
    The blue and green curves correspond to the sky cubes with all components, with no noise added to the visibilities, 
       before (blue) and after(green) foreground subtraction using polynomial fit.  
      The orange curve is the auto-correlation spectrum from the  corresponding  source cube,  its normalisation  is therefore arbitrary. \label{fig:autospec_midlat}} 
\end{figure}

We present on figure \ref{fig:autospec_midlat} the auto-correlation power spectra without noise we obtain after  averaging over frequencies between 1350 and 1375 MHz. The spectrum prior to foreground subtraction is truncated below $\ell\sim 100$ and above $\ell \gtrsim 750$, mainly as a result of the map-making and filtering procedure described  in  section \ref{subsec:pipeline}. In between these two $\ell$ values, the effect of the foreground subtraction (polynomial fit) decreases the  auto-correlation spectrum by 3 to 4 orders of magnitude. For comparison we also show the variations of the auto-correlation power spectrum computed from the 'reference source cube'. 
Due to the  incomplete sky coverage of the survey, \lastchange{the different $\ell$-modes are correlated with each other.} We smoothed the power spectra with a $\Delta\ell=15$ gaussian to mitigate this effect. 

\begin{figure}
        \includegraphics[width=.53\textwidth]{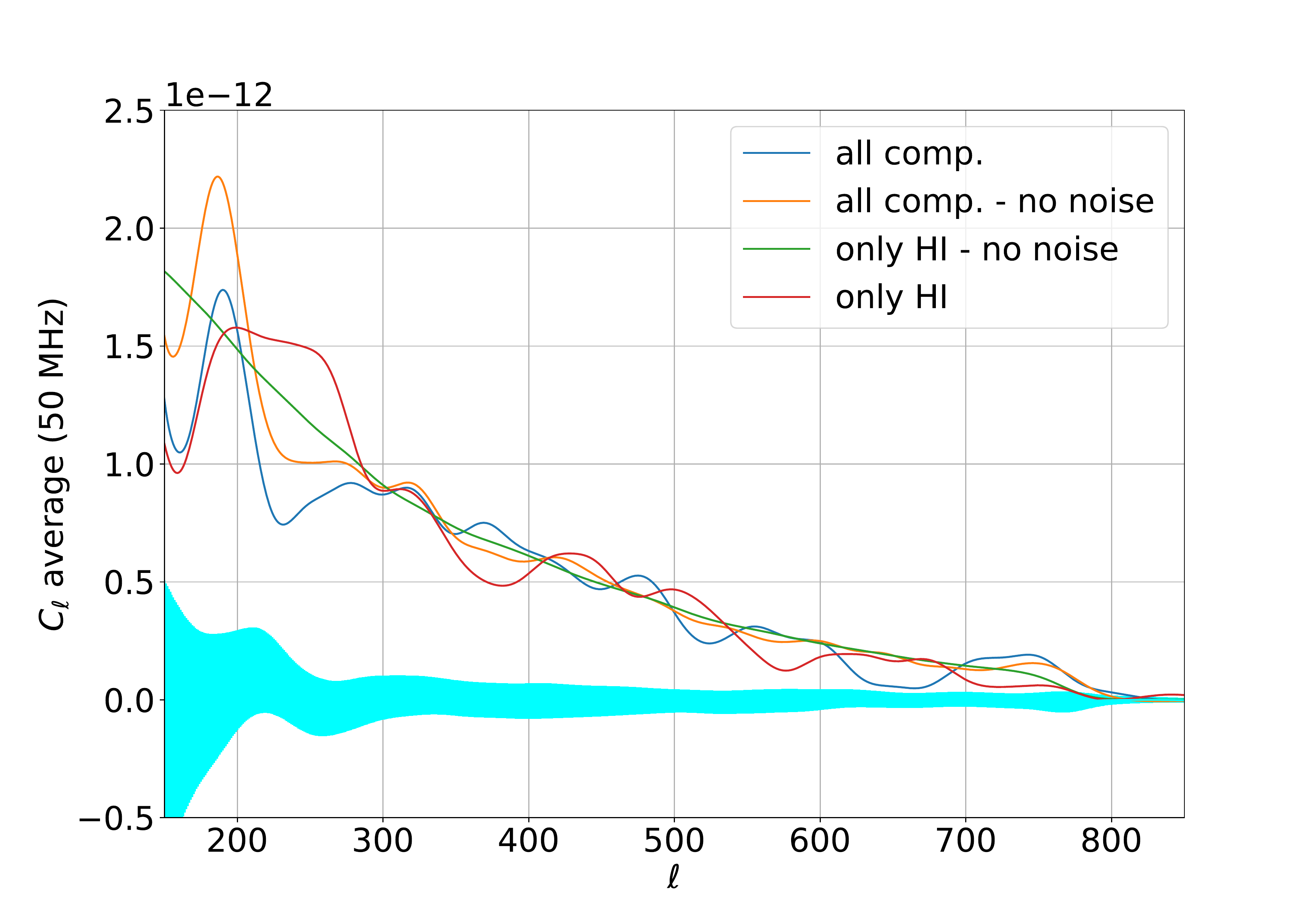}
	\caption{Smoothed cross power \lastchange{spectra averaged over the highest frequency bin (lowest redshift, bin 3), for different simulation cases,} after foreground subtraction. The cyan band outlines the dispersion around central values of the average of the cross power spectra between maps from the simulation combining all components and noise and the 100 shuffled data cubes.\label{fig:spec_midlats_etats_shuf}} 
\end{figure}

We present in figure \ref{fig:spec_midlats_etats_shuf} 
a set of power spectra starting with the ideal situation of a simulation including only \HI sources and without noise, then adding noise and other components, and progressing to the complete simulation of all component with noise. 
\begin{figure}
	\centering
	\includegraphics[width=.5\textwidth]{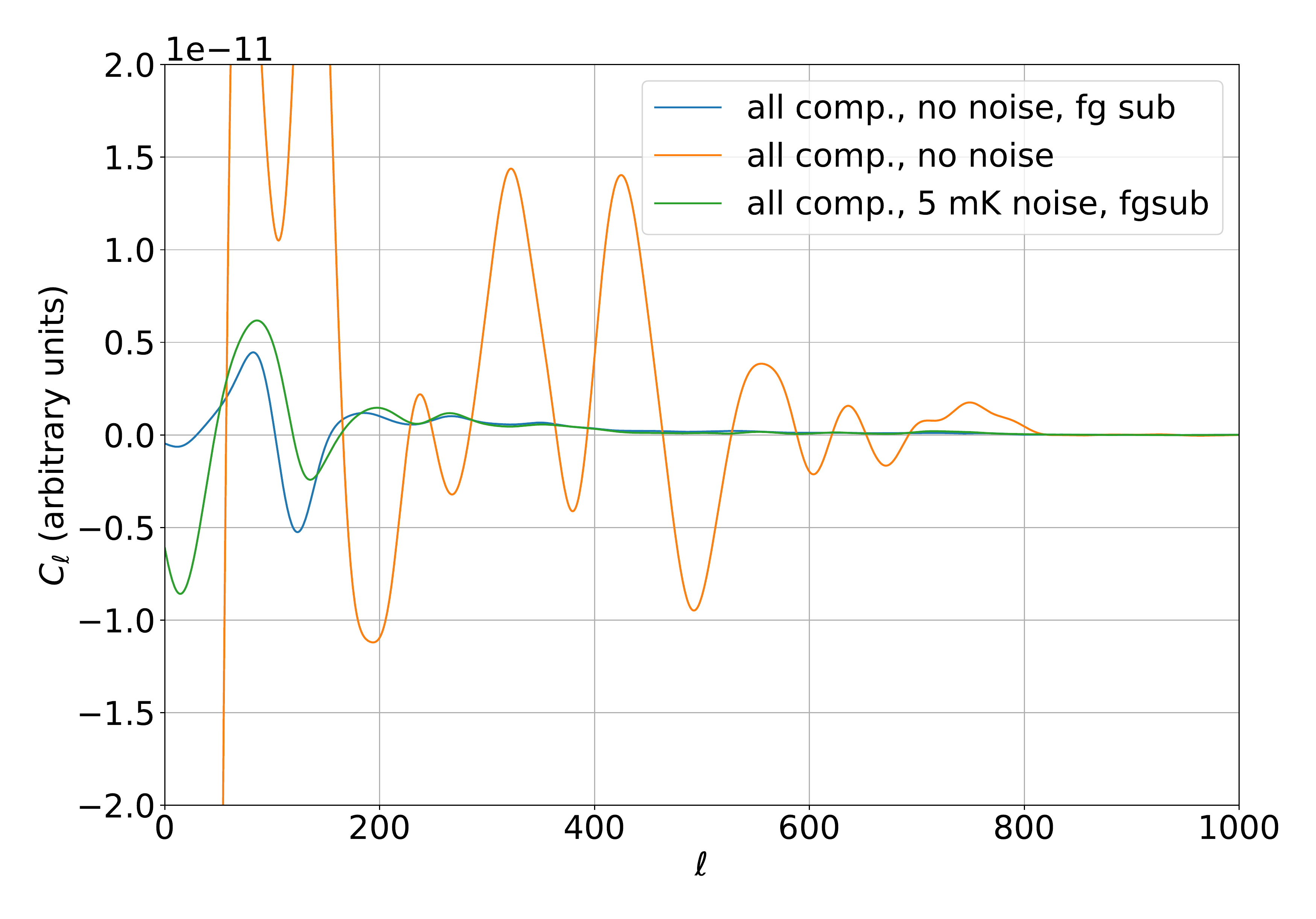}
	\caption{Average of cross-spectra coefficients (\cellcross) from frequencies between 1350 and 1375 MHz. The orange and blue curve represent the cross-spectrum obtained between the 'reference plane' and data simulated with all astrophysical components but no noise, respectively before (orange) and after(blue) foreground subtraction (polynomial fit). The green curve shows the result obtained with 5 mK noise added, after foreground subtraction. All spectra from this figure have been smoothed with a width of 15, to damp $\ell$-to$\ell$ correlations in the raw cross-spectra. 
	\label{fig:Xspec_midlat}} 
\end{figure}

We can observe the impact of adding noise or including more components which 
leaves systematic residuals in the maps after foreground subtraction, due  {\it e.g.} to mode mixing. 
We observe that  these two effects have roughly similar impacts in terms of the cross-power spectra between simulated planes and the data cube, as could be expected from the analysis reported in  section \ref{subsec:noiselev-survey-sens}.
We also note that in a broad $\ell$ range, the averaged cross power spectrum for the complete simulation (in blue) stays positive, and well outside the dispersion from the 100 shuffled cubes. This reinforces the indication that Tianlai could observe a significant cross-correlation with the SDDSS catalog, when performing a mid-latitude, low-$z$ survey.

One might wonder if the foreground subtraction is a necessary step for cross-correlation detection. 
We present in figure \ref{fig:Xspec_midlat} the averaged cross-power spectra between the 'reference planes' or sources planes 
and sky cube reconstructed from visibilities,
for the frequency interval $1350-1375 \, \mathrm{MHz}$ for three cases.  
The blue and orange curves show 
the cross-correlation from visibilities including all astrophysical components but no noise, 
with and without foreground subtraction. The green curve corresponds to the foreground subtracted maps,
computed from visibilities with noise. 
The improvement brought by the foreground subtraction for extracting a significant cross-correlation signal is clearly visible.
The cross-power spectrum amplitudes of the  blue and green curves, after foreground subtraction, unlike the orange curve, 
stay positive for a  multipoles in a broad  range, for example $250\lesssim \ell \lesssim 500 $,  less affected  map-making, filtering and foreground subtraction procedures. 

\begin{figure}
	\centering
		\includegraphics[width=.5\textwidth]{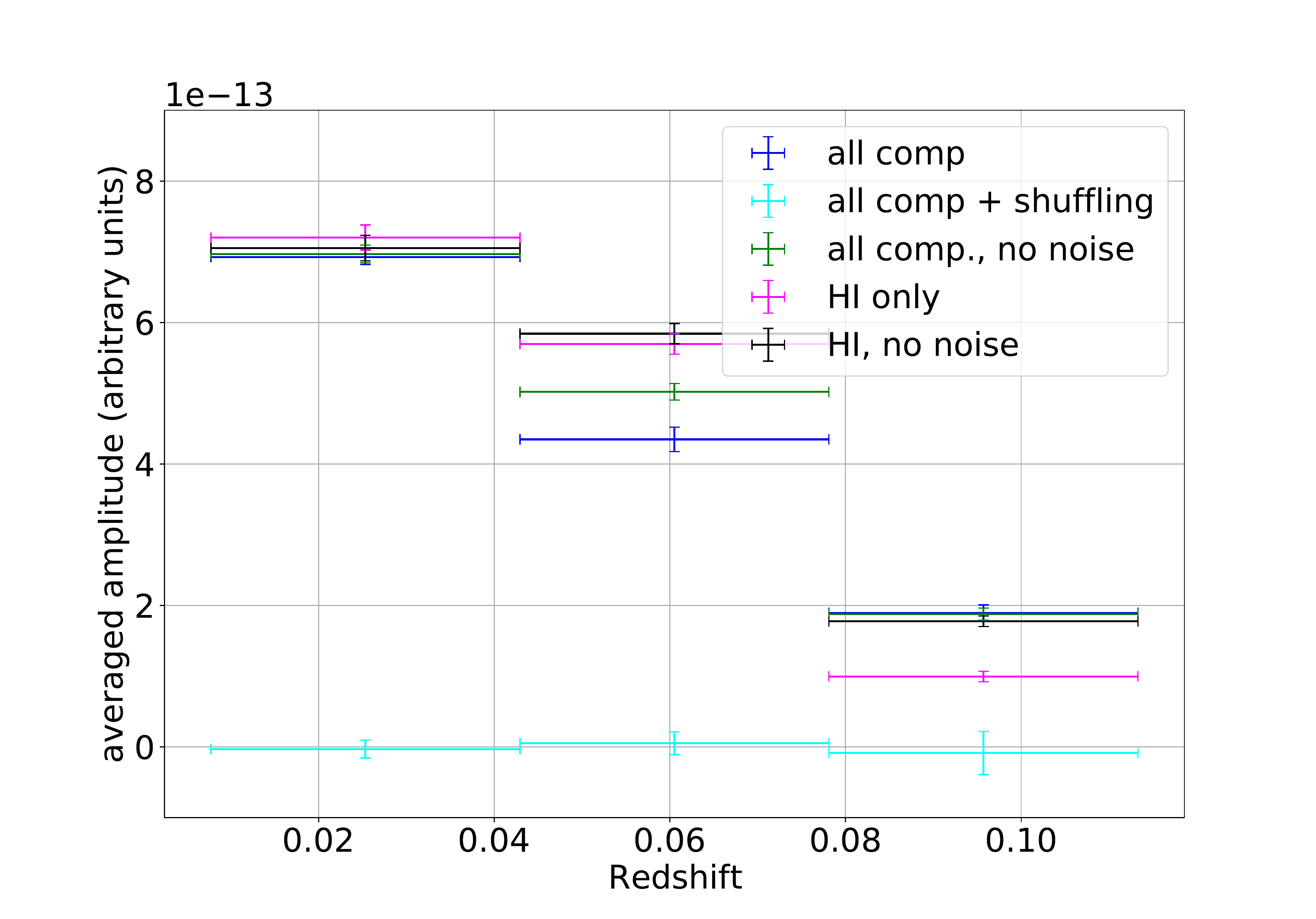}
	\caption{Average of the cross-correlation (\cellcross) in the interval $\ell \in [250,500]$  for each of the redshift bins defined in table \ref{tab:freq_z_bins}. 
	We compare these for different components and noise combinations. All cross-correlation spectra are in arbitrary units.
			\label{fig:cofell_avg_midlat}}
\end{figure}

To summarize the cross-correlation detection perspective, we compute the average of the cross-spectra amplitudes \cellcross for $\ell \in [250,500]$, well \lastchange{outside} the $\ell$ range 
affected by map-making and filtering artefacts.
We present results of this averaging procedure for the three redshift bins on figure \ref{fig:cofell_avg_midlat}. 
\lastchange{We observe that in the two lowest redshift bins the averaged cross-spectra of the simulation case including foregrounds, noise and \HI sources is positive, significantly different from zero,
when compared} to the dispersion computed from the shuffled source cubes.
  These differences amount to $43$ and $ 18$  standard deviations, respectively.
Although this statistical significance estimate is very crude, it still shows that a low-redshift mid-latitude  survey 
operated in the conditions described here  with the Tianlai dish array 
would show a very significant correlation when matched with SDSS low redshift galaxies.

\subsection{Cross-correlation forecast from the NCP survey}

No comprehensive redshift galaxy survey covering the NCP area is available. 
We thus first forecast the sensitivity of a Tianlai survey toward the NCP for 21cm-optical cross-correlation detection 
using a rotated  SDSS catalog, providing an artificial spectroscopic coverage of the north celestial pole.  
In a second step, we use the foreseen characteristics of the ongoing spectroscopic NCP survey to assess the 
fact that it will be shallower than the SDSS. 


\subsubsection{With a rotated SDSS catalog}
\label{sec:sdss_rot_xcor_ncp}
We start with the catalog described in appendix \ref{sec:annex-photo2hi}. 
and we rotate the sky coordinates in order to align $(\alpha, \delta)=(180,45) \deg$ with the NCP, 
thus getting an artificial coverage of the NCP area. 
We use a pipeline similar to that used in the mid-latitude cross-correlaton study presented in section \ref{sec:midlat_xcor} 
with some changes in the last stages. Given the range of declinations  ($83\deg \leq \delta\leq 90 \deg$) studied here, 
we restrict the projection  to a circular area around the NCP, within a radius of 7 degrees. 
The spherical maps are projected into small square flat maps ($169 \times 169 $ maps  with $5 \,  \mathrm{arcmin}$ pixels)  
through a gnomonic projection. A source cube with identical resolution is constructed from the rotated catalog, 
with the same prescription as for the mid latitude case,  with smearing in angular direction, and in frequency,  according to velocity dispersion.  
We also create  100 shuffled source cubes to assess statistical and systematic effects. 
For the NCP analysis we have use the same  frequency intervals as those listed in table \ref{tab:freq_z_bins}. 
Given the small angular extent of the maps, the cross correlation power spectra are computed using a standard Fast Fourier Transform (FFT), rather 
than a spherical harmonics transform. 

\begin{figure}
	\begin{tabular}{cc}
		\includegraphics[width=.5\textwidth]{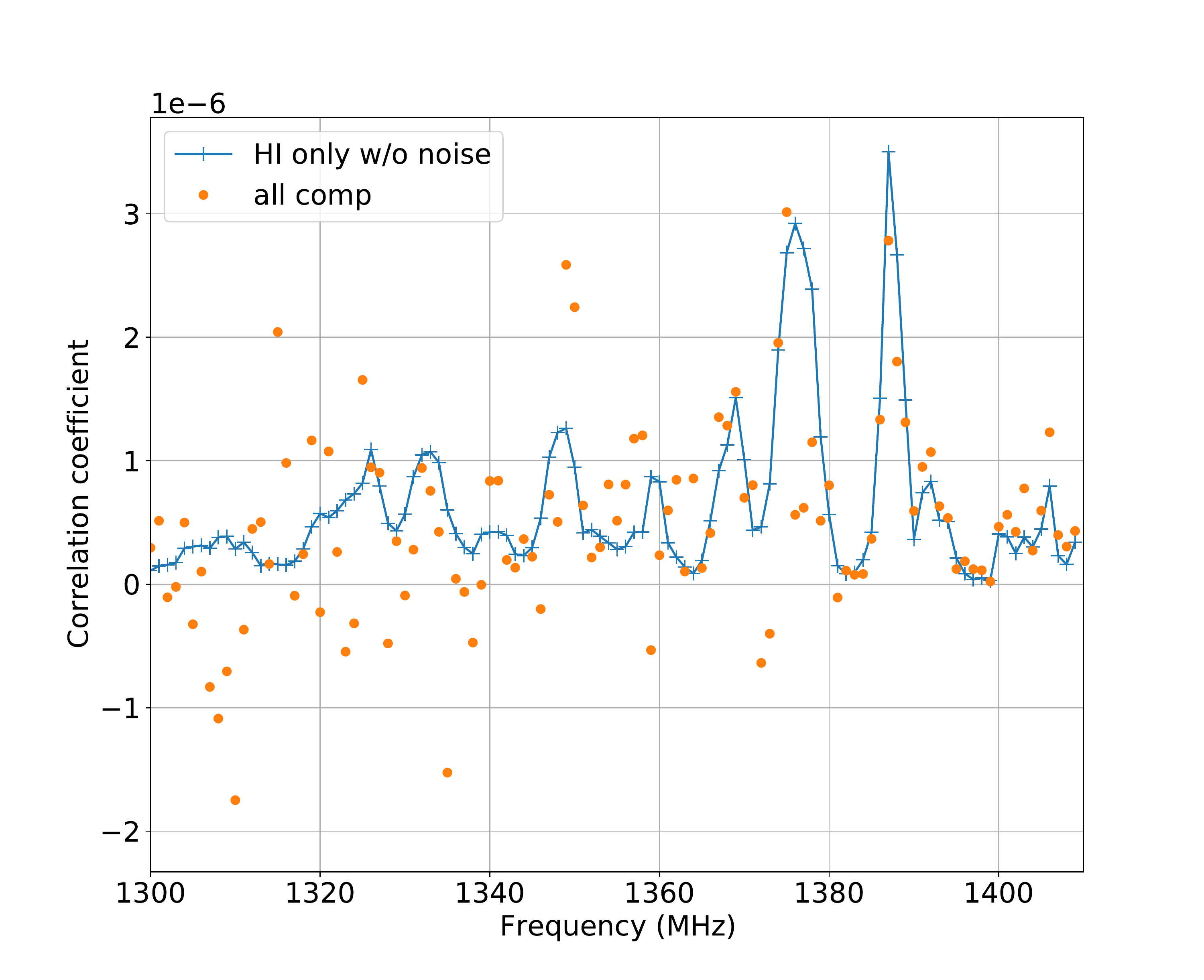}
	\end{tabular}
	\caption{ Raw correlation coefficient computed $\langle \mathrm{RecSky(\nu) \times Src(\nu)} \rangle$ , for each frequency plane. Blue curve 
		(and crosses) correspond to the case where only \HI sources were contributing to the simulated visibilities, 
		while orange circles correspond to the case where visibilities were computed from 
		all sky components (foregrounds and \HI sources) and include $5 \mathrm{mK}$ noise. 
		\label{fig:coef-cor-ncp-freq}}
\end{figure}

As a first test of the presence of a correlation between the simulated sky and the source cube, we 
computed the correlation coefficient the two maps for each frequency plane $\langle \mathrm{RecSky(\nu) \times Src(\nu)} \rangle$.    
Figure \ref{fig:coef-cor-ncp-freq} represents the evolution of these  raw correlation coefficients, as a function of frequency.  Large values of the correlation coefficients computed for the ideal case (no foreground, no noise, in blue), appearing 
as peaks in the distribution, near 1387 MHz or 1370 MHz for example, might be interpreted as the sign of presence of non linear clustering,  
maybe sheets or filaments appearing in the galaxy distribution. Most  features seen above 1370 MHz in the ideal case stay visible in the realistic case where all sky components 
and noise are included (orange circles), for example the peaks around 1375 and  1386 MHz. 

\begin{figure*}
	\begin{center}
		\begin{tabular}{cc}
			\includegraphics[width=.5\textwidth]{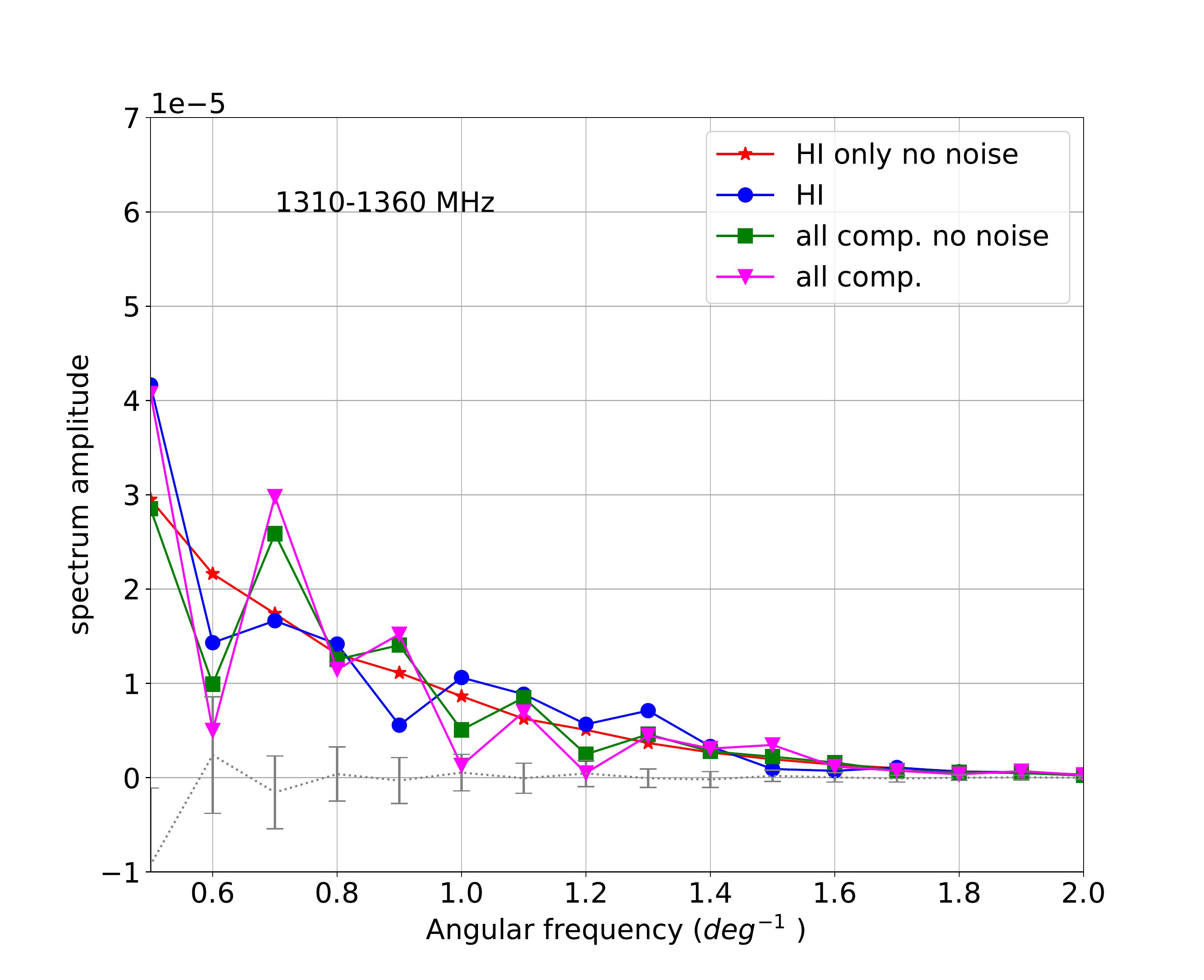}
			&
			\includegraphics[width=.5\textwidth]{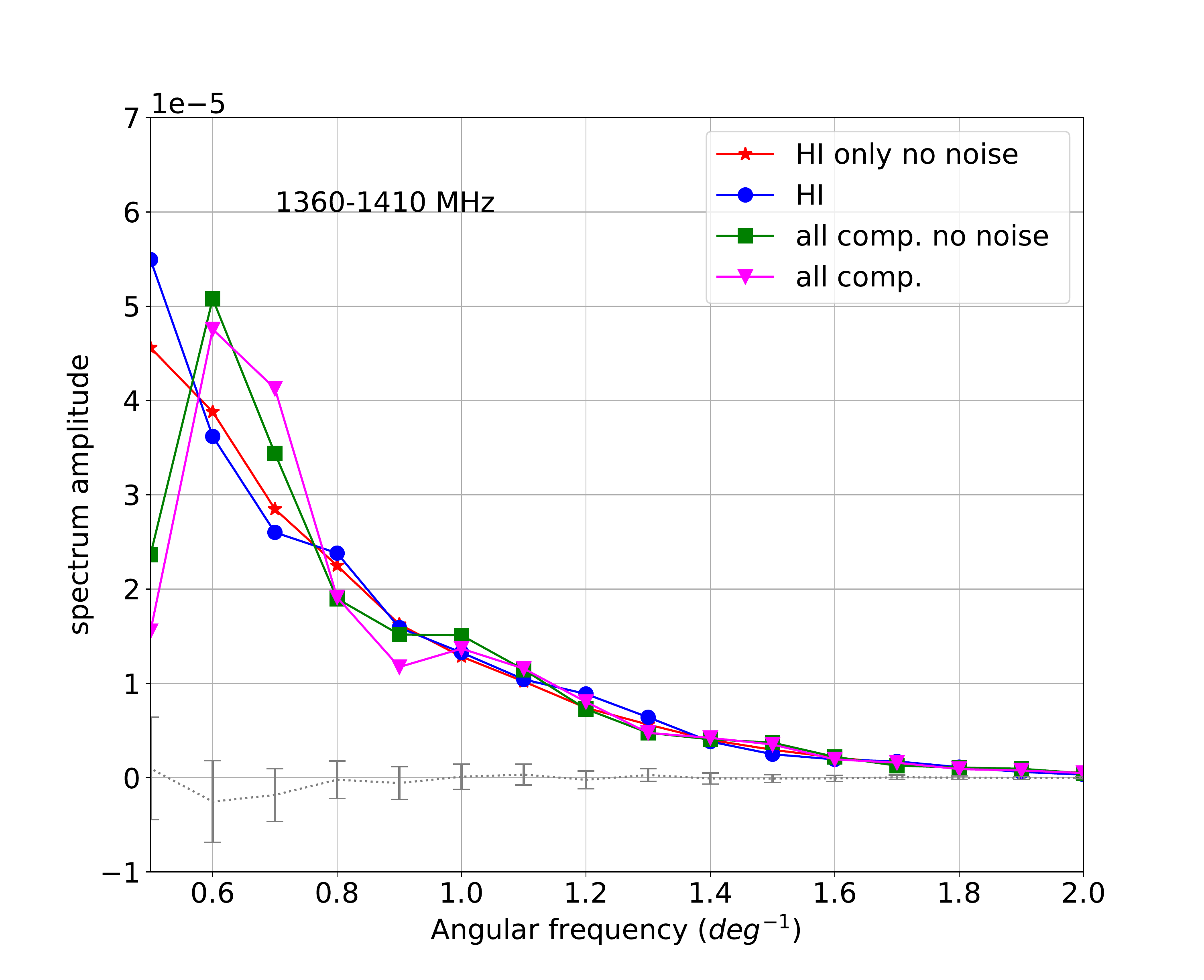}
		\end{tabular}
	\end{center}
	\caption{Reconstructed averaged  cross-spectra in the  two frequency intervals corresponding to bins 2 (left) and 3 (right) from table \ref{tab:freq_z_bins}. We compare results obtained from various simulation parameters, from the ideal (only HI sources, no noise) to the complete  case. On both sides, the cyan results represent  the averages and dispersions from the power spectra from the cross-correlations with a set of 100 shuffled catalogs. As we use only the sources' positions to build their cube, units are arbitrary.  \label{fig:1dspec_ncp_stages}} 
\end{figure*}

To evaluate more thoroughly  the cross-correlation between the 
simulated radio maps and the optical catalog derived maps,
and its angular scale dependence, we compute 2D FFTs for each pair of maps of a given frequency plane. 
We average the amplitudes of the FFT modes in bins of angular wave-modes (azimuthal average), 
which results in  1D amplitude vectors (one per frequency plane). 
We present in figure \ref{fig:1dspec_ncp_stages} averages of such spectra, 
in the second and third  frequency intervals, 
1310-1360 MHz,  on the left panel and 1360-1410 MHz, on the right panel. 
We restrict the range of angular frequencies in these plots to $0.5 - 2 \, \mathrm{deg}^{-1}$ since, on the one hand, 
at low angular frequencies, i.e. large  angular scales, maps are dominated by reconstruction and foreground subtraction artefacts, 
and by noise at high angular frequencies.

Figure  \ref{fig:1dspec_ncp_stages}  presents a comparison of the power spectra reconstructed for different simulation cases 
with the ideal case including only \HI sources and no noise, shown as red stars.
In that case, we get a smooth and positive spectrum which exhibits a very significant positive cross-correlation, as judged from the dispersion 
obtained in the 100 shuffled realisations, shown in gray.
We can observe that adding noise (blue circles) significantly degrades this at large angular scales ($\lesssim 2^\circ$).  
Nevertheless, these spectra amplitudes stay positive, with higher statistical significance in the highest frequency interval (1360-1410 MHz). 
The foreground residuals, observed when including all astrophysical components, 
but without adding noise to visibilities,  
also affect the cross-correlation spectra, shown as the green squares in figure \ref{fig:1dspec_ncp_stages}). 
The impact of the foreground subtraction residuals appears subdominant, compared to that of the noise, in agreement with survey sensitivity analysis presented in  \ref{subsec:noiselev-survey-sens}. 
Finally, with noise added, the cross-spectra (shown as magenta triangles) in both frequency intervals  stay significantly positive. 

Using the dispersions extracted from the shuffled simulations  we can compute the statistical significance 
of  the distance from a null value of obtained cross-power spectra. 
Doing this in each of the 3 frequency intervals of table \ref{tab:freq_z_bins}, and averaging over the 
$0.6-1.3 \mathrm{deg^{-1}}$ angular frequencies, we find a significance of $23.8, 7.5,$ and $5.8$ 
standard deviations for the cross-correlations at
1385 MHz, 1335 MHz and 1285 MHz respectively.  
We can therefore conclude that we would detect with a high significance a cross-correlation between 
an optical galaxy catalog with similar characteristics, completeness and sensitivity, as the SDSS 
covering the region surrounding the NCP, at least for the lowest  redshift interval. 
The NCP cross-correlation is significant, but lower than in our  
forecast for a mid-latitude survey. 
However, we note that, as shown in section \ref{subsec:noiselev-survey-sens}, instrument noise is the main limitation for the NCP survey 
and the noise per visibility sample we have used here is conservative, hence so is our forecast. 
On the other hand, the sky area covered by the NCP survey is much smaller than the mid-latitude case, 
which does  explain part if not all of the difference between the mid-latitude and NCP cross-correlation significance.

\subsubsection{What can we expect with the future Tianlai NCCS-based spectroscopic catalog ?}
\label{sec:nccs-ncp_crosscor}
The Tianlai collaboration is currently  carrying a spectroscopic survey based on the NCCS photometric 
catalog,  performed with the WYIN telescope\footnote{\href{https://www.wiyn.org/}{https://www.wiyn.org/}} 
and its HYDRA spectrograph\footnote{\href{https://www.wiyn.org/Instruments/wiynhydra.html}{https://www.wiyn.org/Instruments/wiynhydra.html}}. A 4 degree radius disk around the NCP have been targeted by the WIYN observations up to now. 
The cleaned version of the NCCS catalog provides, in addition to purely photometric information, an indication on the point-like or extended nature of each object, called the {\tt PESS} score. They advertise that a score greater than 2 indicates an extended source used to select objects for spectsoscopy. 

In order to get a realistic evaluation of the  cross-correlation signal strength which can be achieved with our NCCS-based spectroscopic catalog, we need to determine the completeness level of this sample w.r.t. that of the SDSS sample used in the previous section.  The photometry in NCCS is provided in the Johnson-Cousins system whereas 
in SDSS the ugriz' system is used. According to \href{https://www.sdss.org/dr12/algorithms/sdssubvritransform/}{https://www.sdss.org/dr12/algorithms/sdssubvritransform/}, specifically their Lupton(2005) relations, we find an approximate magnitude conversion equation $R=r-0.25$ using the average color $r-i \simeq 0.35$ of our SDSS galaxies for $z<0.1$.  We compared 
the number of objects in our SDSS rotated catalogs, with $z<0.1$ and $r<17.5$, within 7 degrees of the NCP with the number of extended objects ({\tt PESS} $\geq 2$) in the same area in the NCCS, with $R<17.75$. We find 1131 and 1027 objects in those samples, respectively. We conclude that the object density in the NCP area (7 degree radius) of the two catalogs agree within 10\%. Note that we assume that galaxies further than $z=0.1$ will most probably be flagged as point-like sources in the NCCS. However, not all {\tt PESS} $ \geq 2$ objects from NCCS are successfully observed at WYIN. The current results from our spectroscopic reduction pipeline show that a reliable redshift is obtained for 40\% of these objects. 

\begin{figure}
		\includegraphics[width=.5\textwidth]{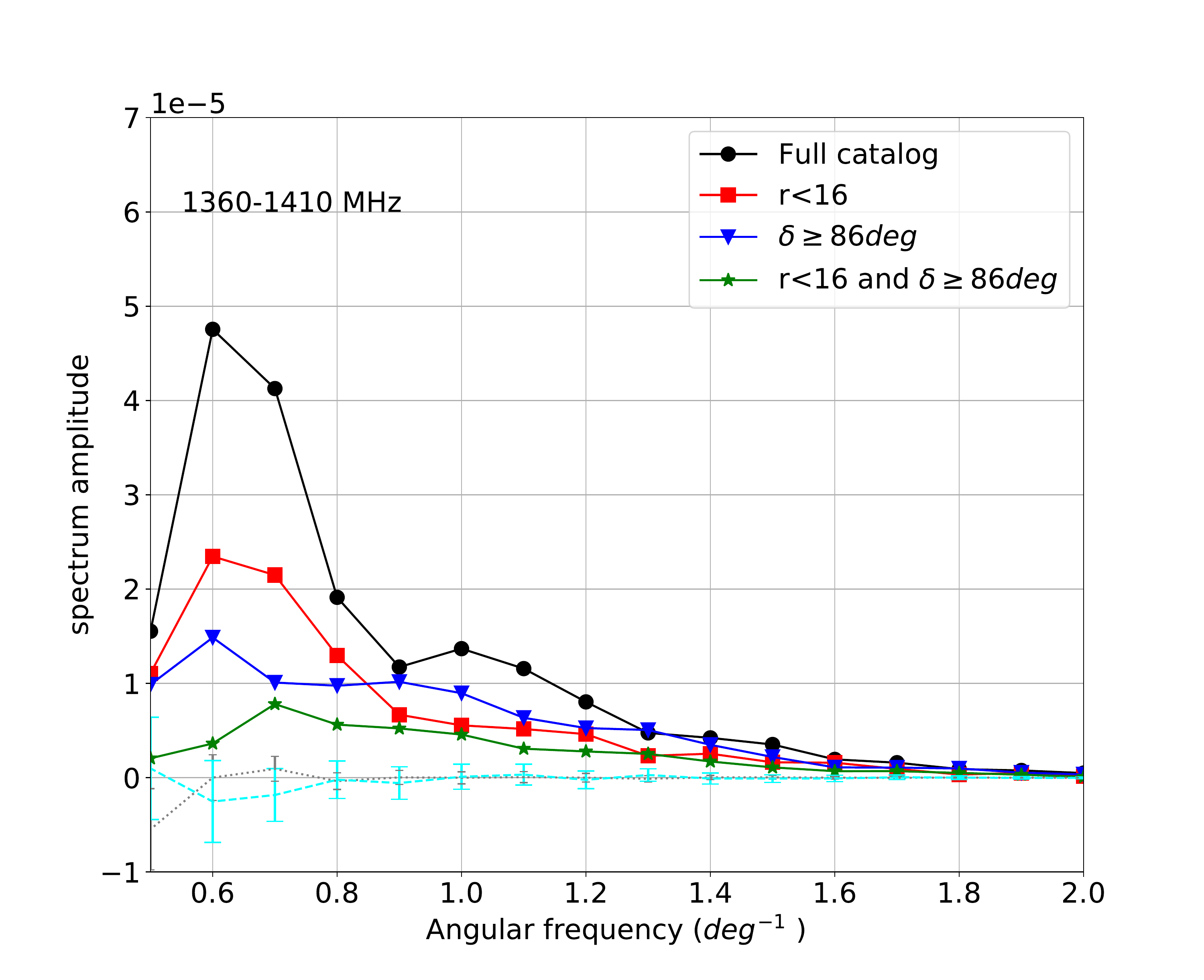}
	\caption{\ra{(caption updated)} Effect of excluding a fraction of optical galaxies when computing the cross correlation for the NCP case for the $3^{rd}$ frequency interval listed in table \ref{tab:freq_z_bins}, i.e. the lowest redshift bin. 
	The 1D cross power spectra as a function of the angular frequency are shown,
	as black circles, when all sources are included, as red squares when only sources with $r<16$ are included,
	as blue triangles for sources with $\delta \geq 86 \deg$, and as green stars, when requiring both 
	$r<16$ and $\delta \geq 86 \deg$. The spectra obtained from the shuffled source cubes, are also shown,
	in cyan (larger ticks) for the larger area, in gray (smaller ticks)for the smaller area.
	The vertical axis is in arbitrary units as the source cube is built from optical object positions only.\label{fig:1dspec_ncp_cut}} 
\end{figure}

Although this efficiency might be increased in the future, we will use this incompleteness factor, translated into a $r$ magnitude cut in the SDSS catalog. 
Looking at the SDSS $r$ magnitude distribution, we find that requiring $r\leq 16$ will result in a similar galaxy count over the NCP area as the one in the NCCS spectroscopic catalog. We therefore estimated the cross-correlation signal level between our simulated maps and the corresponding 
optical data cubes built while imposing $r\leq 16$, and compared these with the estimation obtained with the full optical catalog. We also performed this analysis using the central 4 degree radius disk only, which is the area currently covered by the on-going WYIN observations. We show the cross-correlation power spectra we reconstruct in these 4 cases, for the highest frequency interval (lower redshift)  in figure \ref{fig:1dspec_ncp_cut}. 

Following the same method outlined in section \ref{sec:sdss_rot_xcor_ncp} we compute the statistical significance of the cross correlation spectra using the dispersion of the null spectra obained from the shuffled cubes. 
We report these results in table \ref{tab:results_ncp_cuts} for  the two highest frequency intervals. Any selection w.r.t. the full sample changes  the signification of the cross-correlation signal, to various extents. From the values reported in table \ref{tab:results_ncp_cuts} we  observe  that the brightness cut has a larger effect for the highest redshift bin :  as the sources' brightness decrease with redshift, the fraction of those rejected by the cut gets larger with redshift. We also note that selecting a narrower region around the NCP has a larger effect for the lowest redshift bin. We may  benefit of the lower noise per map pixel in this central area. 
Also, as the solid angle included in the analysis increases with redshift, more sources are included in the optical catalog at higher redshift. 

\begin{table}
    \caption{Number of standard deviations of the cross-correlation between the optical data cubes and the simulated observations for various  optical seections, in the two highest frequency intervals in our analysis. \label{tab:results_ncp_cuts}}  
    \centering
    \begin{tabular}{c|c|c|c|c}
    Frequency range  & all & $r\leq 16$ & $\delta\geq 86\deg $ &   $r\leq 16$ and \\
     & & & & $\delta\geq 86\deg $ \\
    \hline
    1360-1410 MHz & 23.7 & 19.3 & 17.5 & 14.7 \\
    1310-1360 MHz & 7.6 & 5.3 & 16.6 & 14.7 \\

    \end{tabular}

\end{table}

We conclude that given the efficiency and sky coverage of our ongoing NCCS spectroscopic survey, a significant cross-correlation between the Tianlai radio observations and the enriched NCCS catalog should be observed, with  more than 14 Std.Dev statistical significance, in the two highest frequency intervals. However, this signal will reach a higher statistical signification if the area covered by our spectroscopic observations can be extendend beyond $\delta \geq 86 \deg$. This would also be true if we succeed in increasing our efficiency at getting redshifts for fainter sources.
Finally, the cross-correlation signal  will be stronger if we achieve a noise per visibility sample lower than the conservative 
value of 5 mK used in our simulation, which is feasible by increasing the observation time.

\section{Conclusions, perspectives}
\label{sec:discussion}
We have shown that low-z surveys carried out by the Tianlai dish array, by tuning the analog electronic 
to the 1300-1400 MHz band would be sensitive enough to see extra galactic 21 cm signal, 
either through direct detection of a few nearby massive \HI galaxies or in cross correlation with optical surveys. 

The Tianlai instrument is designed to observe in transit mode, with sky coverage obtained through constant declination scans. 
By pointing the antennae toward the North Celestial Pole (NCP), a small sky area is covered, leading to increased sensitivity thanks 
to long integration time. We have thus studied two complementary survey strategies in this paper, a deep survey covering 
$\sim 150 \mathrm{deg^2}$  around the NCP, and a shallower survey, corresponding to  a $\sim 10 \mathrm{deg}$ declination 
band at mid-latitudes, covering $\sim 1500 \mathrm{deg^2}$ useful area around $\delta = 55^\circ$ declination 
and overlapping the SDSS footprint.
 
A noise level of $\lesssim 2-4 \mathrm{mK}$ per $1 \mathrm{MHz} \times 0.25^2 \mathrm{deg^2}$ pixels, 
should be reached for a 3 months survey of 
the NCP area,  leading to a point source detection threshold of $\sim 0.08 \mathrm{Jy}$. Our study shows that the residuals 
from the mode mixing for the NCP survey should be negligible compared to the noise fluctuations.  Extending the survey duration 
to a year would thus decrease the noise to $\sim 1-2  \mathrm{mK/pixel}$ and reach a source detection threshold of $\sim 0.05 \mathrm{Jy}$. 
Tianlai should then be able to detect  $\gtrsim 10$ nearby \HI clumps or galaxies through their 21cm emission, up to redshifts 
$z \lesssim 0.02$.  
We have also studied the possibility to detect statistically the extragalactic 21cm emission, dominated by the one from galaxies 
at these low redshifts, in cross correlation with optical surveys. We show that the cross-correlation 
signal between Tianlai NCP intensity maps and the NCCS optical galaxy survey could be detected with very high significance, $\sim 15$ standard deviations,
provided we get redshifts for most of the NCCS galaxies with magnitude \lastchange{$R<16.25$}. 
A spectroscopic survey of NCCS galaxies is indeed being carried out with the WIYN telescope, and 
it would be helpfull  to pursue this effort to extend to the surveyed area 
to the full 7 deg. radius disk  foreseen for the planned Tianlai low-z NCP survey.

We have adopted a conservative approach throughout this paper, using the $5 \mathrm{mK}$ noise level for visibilities, corresponding 
to a 3 months survey and not the $2.5 \mathrm{mK}$ value corresponding to the nominal 1-year NCP survey. This leaves some room 
to mitigate a fraction of increased fluctuations associated with calibration errors, partial beam knowledge and possibly correlated noise.
Further studies are needed to assess the impact of these instrument imperfections on the survey performance. As an example, our preliminary 
investigation shows that a $7 \mathrm{deg}$ RMS phase calibration errors between baselines would lead to a 3-fold  increase of the residual 
fluctuations at the map level for the NCP survey. 

The mid-latitude survey suffers larger residuals from mode mixing, as well as higher noise level due to the larger sky area,
but has the advantage of having a large overlap with the SDSS spectroscopic survey at low redshifts.  
The lower residuals due to mode mixing (frequency dependent synthetised beam) for the NCP case can be explained by 
the larger number of effective baselines created by the rotated array, projected on the same sky area due to the earth 
rotation, while this rotation makes different parts of the sky to drift in front of the instrument at mid-latitudes. 
The fluctuations due to the noise reaches about $\sim 15 \mathrm{mK/pixel}$ for the mid-latitude survey, with residuals 
from foreground subtraction at a similar level, leading to a direct source detection threshold about $0.9 \mathrm{Jy}$, 
about ten times higher than that of the NCP case. Despite this significantly higher level of residuals and noise in the foreground 
subtracted maps, our study shows that the cross correlation of the Tianlai foreground subtracted maps with the 
SDSS optical galaxies could be detected with high significance ($\gtrsim 40 \, \mathrm{Std.Dev}$).  

The results presented in this paper can be consolidated in the future by studying the impact of instrumental effects not yet taken 
into account, phase and amplitude calibration errors and drifts, correlated noise, as well as the impact of poorly 
known individual antenna beams, specially far side lobes. There is also room for optimising further the observation strategy, 
for example, the question of the optimal area to be covered in the NCP region for a given survey duration. Another important 
parameter would be the exact frequency coverage. However, Tianlai already has three sets of filters, covering 1170--1270, 1250--1350 MHz and 1330--1430 MHz, respectively, in addition to the currently used 700--800MHz filters.  
The contribution of different frequency slices vary due to the decrease of the sensitivity with the redshift, 
partially compensated by the volume increasing with the redshift. 
One has also to take into account the frequency intervals which would be impacted 
by RFI to determine the optimal frequency band.  
 
\appendix
\section{Simulation of extragalactic \HI 21cm signal}
\subsection{Preparation of the input catalog from SDSS}
\label{sec:annex-sdss}
We extract SDSS data through their SQL server with a geometric selection of the intersection of the {\tt PhotoObj}, {\t tSpecObj} and {\tt Photoz}
 catalogs (the latter providing absolute magnitudes). From this initial catalog we select objects satisfying the empirical fiducial criteria: 
 \begin{itemize}
 	\item sources belonging to the category {\tt GALAXY}
 	\item  a spectroscopic redshift in the interval [.005,1.0]
 	\item selection of objects with ordinary colors: $-0.5<r-i<2.5$ and $-0.2<g-i<1.65$ to avoid a few rare outliers 
 \end{itemize}

\subsection{From optical photometry to radio parameters}
\label{sec:annex-photo2hi}

In order to be able to simulate observations of a sky composed of diffuse and point-like continuum radio sources as well as 21cm emission from galaxies, we need to determine 
the 21 cm emission properties of the latter starting from an optical galaxy catalog. We follow a two-step procedure to achieve this. First, following \cite{galactMstar} we estimate the stellar mass of each galaxy from their photometric properties using their equation 8\footnote{We use this equation as found in the MNRAS published version of \cite{galactMstar}, where it seems to have been corrected  w.r.t. the arXiv preprint.} : 
\begin{equation}
\log( M_\star/M_\odot) = 1.15+0.70(M_g-M_i)-0.4M_i
\label{eq:mstar}
\end{equation} 
This equation can easily be applied for SDSS objects, using their Photoz table that includes absolute magnitudes. 
From a cross-match between the ALFALFA and SDSS, \cite{alfalfa_mstar} investigated the relation between the \HI and stellar masses of galaxies. From their figure 3, where results of stacking ALFALFA observations for the full SDSS sample  are reported, 
we extract a simple relation between the \HI and stellar masses : 
\begin{equation}
\log(M_{HI}/M_\star) = 0.179 - 0.66(\log M_\star - 9.21)
\label{eq:mhi}
\end{equation}
Combining equations \ref{eq:mstar} and \ref{eq:mhi} we can now estimate the \HI mass for each galaxy. \cite{width_himas_alfalfa} (their figure 7) show 
the relation between the \HI mass and $W_{50}$, the gas velocity dispersion width  at half maximum flux, from which we infer a simple linear relation between $\log W_{50}$ and $\log M_{HI}$.

Using this equation and \ref{eq:s21_mhi} 
we can finally  estimate the 21cm peak flux and frequency width for each galaxy, after accounting for their distance 
as determined by their redshift within a fiducial cosmology (Planck $\Lambda$CDM). 
  
\begin{figure*}
	\centering
	\begin{tabular}{cc}
		\includegraphics[width=.55\textwidth]{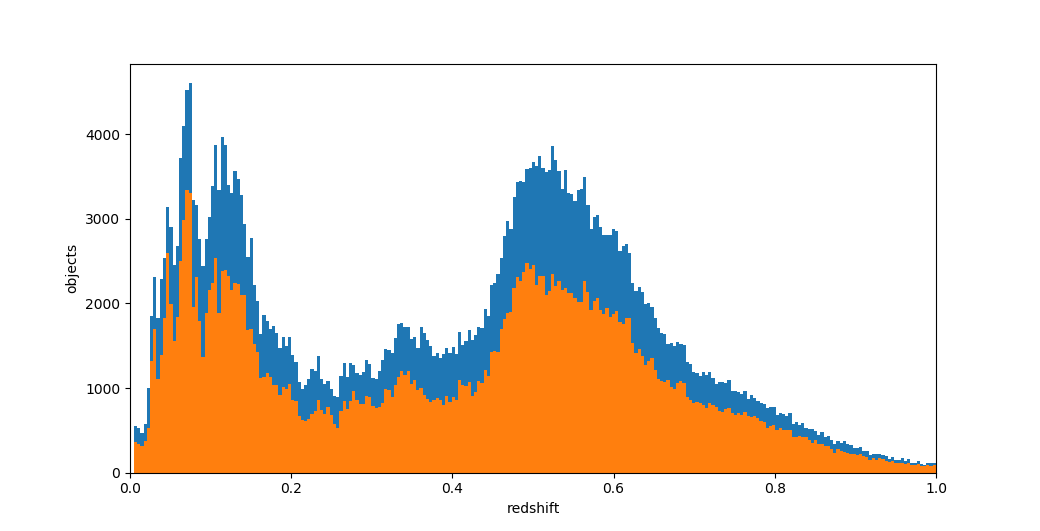}
		&
		\includegraphics[width=.4\textwidth]{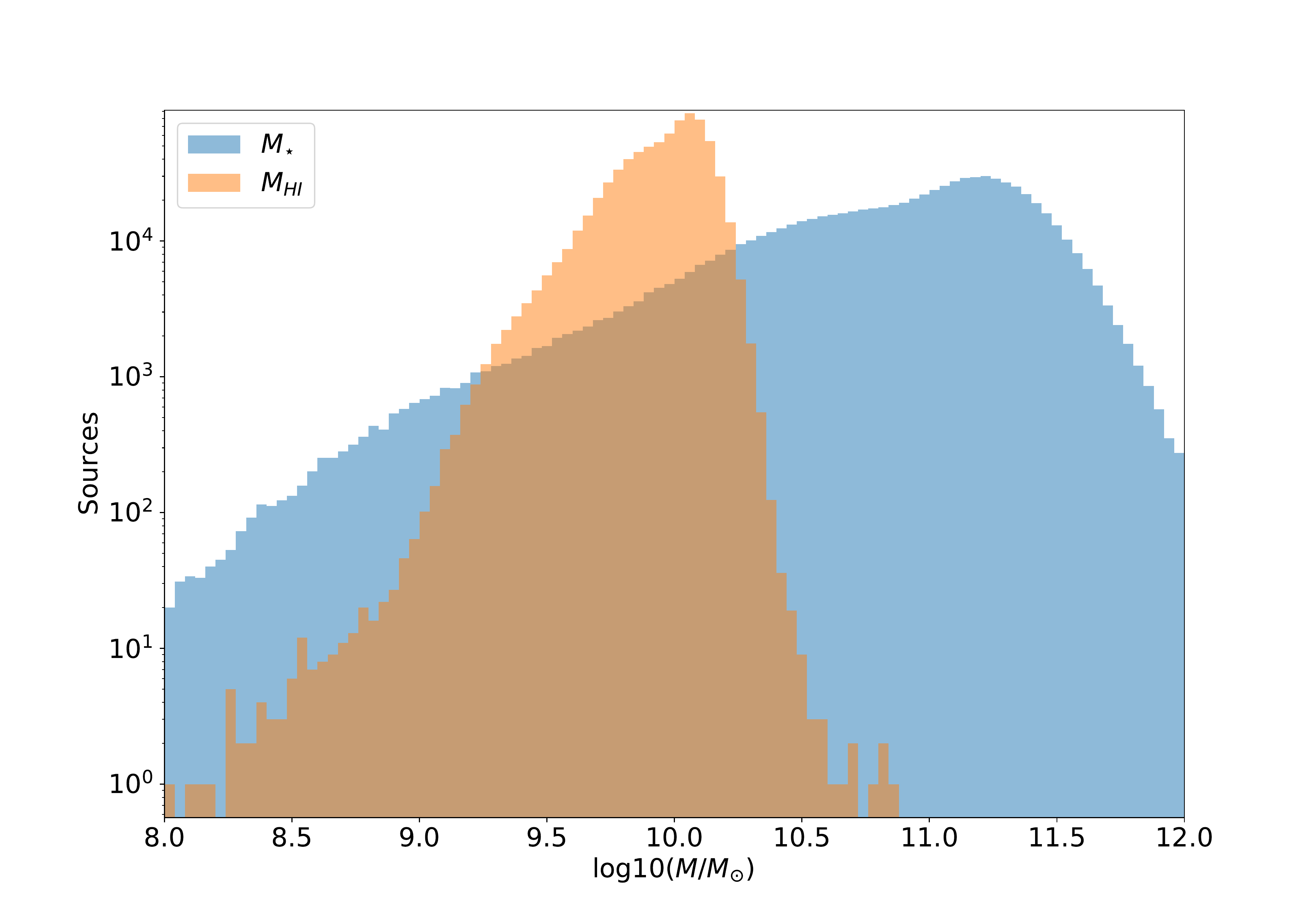} 
	\end{tabular} 	
	\caption{Left panel: Redshift distribution of the SDSS sources (blue histogram) and after additional selection criteria described in paragraph \ref{sec:annex-sdss}, in orange. Right panel:  Derived stellar mass distribution in blue, and corresponding \HI mass distribution in orange. \label{fig:massehi_z}}  
\end{figure*}

\begin{figure*}
	\centering
	\begin{tabular}{cc}
		\includegraphics[width=.45\textwidth]{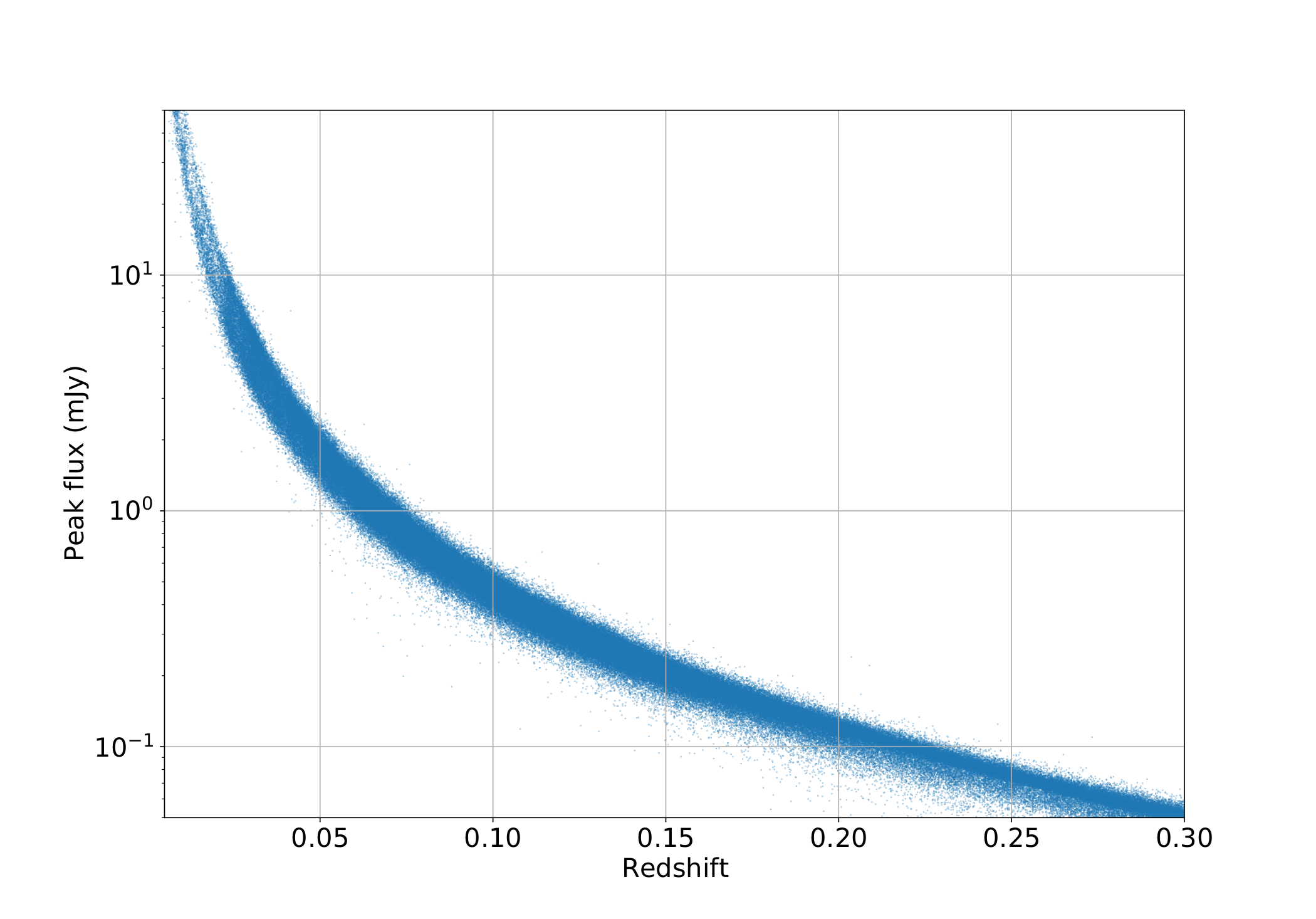}
		&
		\includegraphics[width=.45\textwidth]{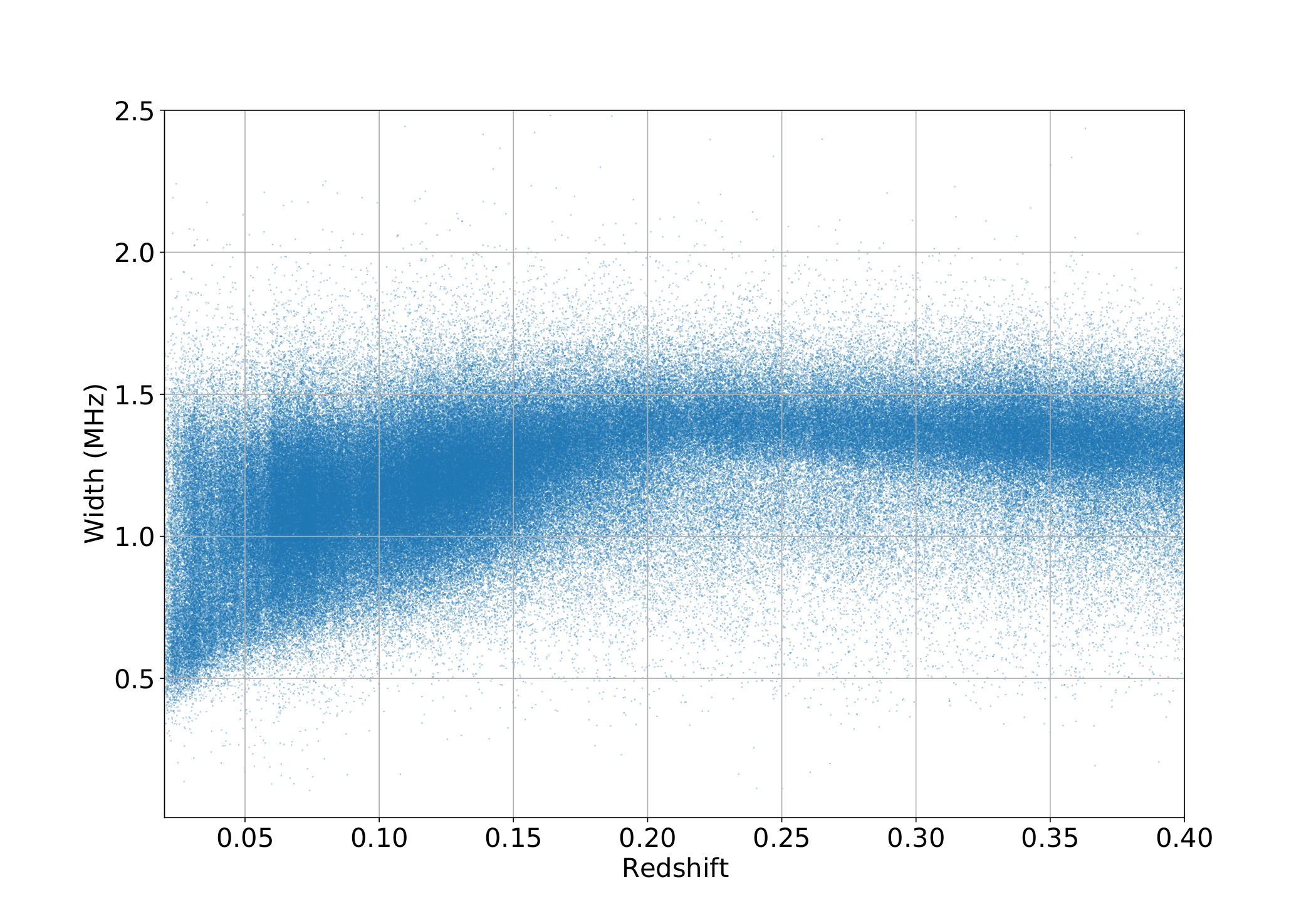} 
	\end{tabular} 	
	
	\caption{21 cm peak flux (in $\mathrm{mJy}$) and frequency dispersion width (in MHz) computed from $W_{50}$ as a function of redshift for our simulated \HI source catalog, derived from the SDSS catalogs.\label{fig:hi_flux_z}}  
\end{figure*}

We prepared a input catalog  for our simulations from the SDSS photo- and spectroscopic catalog,  choosing galaxies with declination above $\delta=30\deg$.
The redshift and the derived stellar and \HI\ mass distributions for these sources are shown in figure \ref{fig:massehi_z}. The computed \HI\ flux and frequency dispersion width 
as a function of redshift are shown in figure \ref{fig:hi_flux_z}.

\section*{Acknowledgements}
This research made use of Photutils, an Astropy package for
detection and photometry of astronomical sources (\cite{photutils_121}).
\section*{Data Availability}
The simulated data generated for and analysed in  this article will be shared on reasonable request to the corresponding author. 
Public datasets used in this article may be retrived from the   references indicated in the text, namely :  \href{https://skyserver.sdss.org/dr16/en/tools/search/sql.aspx}{SDSS DR16 server}, 
 \href{https://lambda.gsfc.nasa.gov/ }{Lambda archive} and \href{https://vizier.u-strasbg.fr/viz-bin/VizieR}{Vizier at CDS}.
\bibliographystyle{mnras}
\bibliography{refs} 

\bsp	
\label{lastpage}
\end{document}